\documentclass[aps,prl,twocolumn,superscriptaddress]{revtex4-1}

\usepackage{graphicx,color}% Include figure files
\usepackage{epsfig}
\usepackage{dcolumn}% Align table columns on decimal point
\usepackage{amsmath,amssymb,bm}
\usepackage{amsmath}
\usepackage{hyperref,url}
\usepackage[noabbrev]{cleveref}
\usepackage{CJK}
\usepackage{lineno}
\usepackage[english]{babel}
\usepackage[T1]{fontenc}
\usepackage{pifont}
\usepackage{amsfonts}
\usepackage{wasysym}
\usepackage{amsbsy}
\usepackage{psfrag}
\usepackage{color}
\usepackage{multirow}
\usepackage{cases}
\usepackage{stackengine}
\usepackage{float}
\usepackage{blkarray}
\usepackage{cancel}
\usepackage{mathrsfs}
\usepackage{empheq}
\usepackage{orcidlink}
\addto\captionsenglish{}
\renewcommand\thefigure{{\textbf{\arabic{figure}}}}

\addto\captionsenglish{}
\renewcommand\thetable{{\textbf{\arabic{table}}}}

\DeclareMathAlphabet{\mathbfsf}{\encodingdefault}{\sfdefault}{bx}{sl}

\providecommand\bnabla{\boldsymbol{\nabla}}
\providecommand\bcdot{\boldsymbol{\cdot}}

\newcommand{\boldm}[1]{\boldsymbol{#1}}
\newcommand{\Ca}{\mbox{\text{Ca}}} 		% Capillary
\hypersetup{
    colorlinks=true,
    linkcolor=blue,
    citecolor=blue,
    urlcolor=blue,
}

\usepackage{fancyhdr}
\fancypagestyle{titlepage}{
\fancyhead[C]{Morphodynamics of surface-attached active drops}
\fancyhead[L]{}
}
\begin{document}

%\linenumbers

%\preprint{}

\title{Morphodynamics of surface-attached active drops}


% \email[]{email}
% \homepage[]{Your web page}
% \thanks{}
% \altaffiliation{}

\author{Alejandro Mart\'inez-Calvo\,\orcidlink{0000-0002-2109-8145}\,}
\email{amcalvo@princeton.edu}
\affiliation{Princeton Center for Theoretical Science, Princeton University, New Jersey, 08544, USA}
\affiliation{Department of Chemical and Biological Engineering, Princeton University, New Jersey, 08544, USA}
\affiliation{Department of Physics, Princeton University, Princeton, NJ 08544, USA}

\author{Sujit S. Datta\,\orcidlink{0000-0003-2400-1561}\,}
\email{ssdatta@caltech.edu}
\affiliation{Division of Chemistry and Chemical Engineering, California Institute of Technology, Pasadena, CA 91125, USA}
\affiliation{Department of Chemical and Biological Engineering, Princeton University, New Jersey, 08544, USA}


\begin{abstract}
Many biological and synthetic systems are suspensions of oriented, actively-moving components. Unlike in passive suspensions, the interplay between orientational order, active flows, and interactions with boundaries gives rise to fascinating new phenomena in such active suspensions. Here, we examine the paradigmatic example of a surface-attached drop of an active suspension (an "active drop"), which has so far only been studied in the idealized limit of thin drops. We find that such surface-attached active drops can exhibit a wide array of stable steady-state shapes and internal flows that are far richer than those documented previously, depending on boundary conditions and the strength of active stresses. Our analysis uncovers quantitative principles to predict and even rationally control the conditions under which these different states arise---yielding design principles for next-generation active materials.
\end{abstract}

%Many living and synthetic active systems resemble liquid drops, where a suspension of active units acquires orientational order and spontaneously generates flow capable of deforming the confining liquid interface. This change in shape, in turn, influences both the self-generated flow and orientational order. While the interplay between flow, order, and interfacial morphodynamics is key to the function of active systems, it remains largely underexplored. Here, we study a continuum model to uncover the morphodynamics of a surface-attached active drop containing a uniform suspension of active units that acquire nematic order. We find that the drop achieves a rich array of stable, steady-state shapes and self-generated flows, which can be reversibly controlled by tuning the anchoring of the active units with the boundaries. This diversity of shapes and flows contrasts with previous works that describe such active drops using the thin-film approximation. We hope that our results inspire and inform experiments in active matter, spanning from living systems like bacterial suspensions and biofilms to synthetic materials such as biopolymer and microtubule suspensions.

\date{\today}

%\pacs{47.20-k 47.20.Ma 47.15.G- 47.55.D- \textcolor{red}{check}}

\maketitle

%%%%%%%%%%%%%%%%% SECTION: Introduction %%%%%%%%%%%%%%%%%%%

%These units can self-organize acquiring orientational order and exert stresses that induce spontaneous flows

Active matter refers to systems whose components consume energy from their environment and convert it into motion and/or biomass, thereby driving them far from equilibrium~\citep{Marchetti2013,zhang2021autonomous,hallatschek2023proliferating}. Many biological and synthetic active systems are compartmentalized and therefore resemble liquid drops, with an interface that separates a suspension of active units from the surrounding environment. These units can self-organize and acquire orientational order, exert stresses, and induce spontaneous flows that can deform the confining interface. In turn, these changes in shape can influence the organization of the active units. This interplay between order, flow, and shape plays a key role in the function of diverse forms of active matter, such as organoids~\citep{fernandez2021surface,ishihara2023topological}, bacterial and amoeba colonies~\citep{beroz2018verticalization,Pearce2019,hartmann2019emergence,Qin2020,nijjer2021mechanical,nijjer2023biofilms,ford2024pattern}, the mitotic spindle~\citep{brugues2014physical,oriola2018physics,oriola2020active}, microtubule vesicles~\citep{keber2014topology}, droplets in active suspensions~\citep{adkins2022dynamics,tayar2023controlling}, phase-separated fibril drops~\citep{fu2024supramolecular}, intracellular and developmental flows~\citep{wensink2012meso,khuc2015cortical,saintillan2018extensile,Needleman2019stormy}, and colloidal and ferrofluid droplets~\citep{khalil2014active,aggarwal2023activity,aggarwal2023thermocapillary}. 

In both natural and synthetic contexts, active drops do not typically exist in isolation, but are attached to surfaces, such as in microbial colonies~\citep{beroz2018verticalization,Pearce2019,hartmann2019emergence,nijjer2021mechanical,nijjer2023biofilms,ford2024pattern}, single-cell and tissue migration~\citep{Tjhung2012,Tjhung2015,perez2019active}, and active wetting phenomena~\citep{khalil2014active,alert2018role,perez2019active,aggarwal2023activity,aggarwal2023thermocapillary,liese2023chemically}. Understanding the morphodynamics of these surface-attached active systems is of fundamental interest in active matter physics and has key implications for both biology and materials science. How does the coupling between orientational order, flow, and interactions with confining interfaces influence the shape of surface-attached active drops? A comprehensive answer to this question is still missing, despite the critical role of this interplay in the functioning of active systems. In particular, while current theoretical models provide useful intuition, they rely on the thin-film approximation~\citep{ben2001fingering,joanny2012drop,Loisy2019PRL,trinschek2020thin,Loisy2020,shankar2022optimal,ioratim2022nonlinear,ford2024pattern}, which assumes that all variables describing the drop vary less in the direction parallel to the
substrate than in the direction normal to it~\citep{oron1997long,craster2009dynamics}. This strong approximation is not applicable to many real-world systems, for which the drop shape, ordering of active units, and self-generated flows can vary in all directions. As a result, current understanding is incomplete. Indeed, by relaxing the thin-film approximation, here, we show that the morphodynamics of surface-attached drops are far richer than was previously known.

%Thus, understanding the morphodynamics of active drops is of fundamental interest in active matter physics and carries significant implications for biology and material science~\citep{needleman2017active,zhang2021autonomous,adkins2022dynamics,tayar2023controlling}.

% which makes them to be far from equilibrium~\citep{Marchetti2013,hallatschek2023proliferating}

%e.g., synthetic biopolymer assemblies, as well as living systems like suspensions of bacteria, microtubules, and actin fibers

%To address this gap in knowledge, here we consider a minimal continuum model that describes an active drop anchored to a rigid, impermeable flat substrate, containing a suspension of active units that exhibit nematic order and spontaneously self-generate flow. The dynamics of surface-attached active drops have been extensively studied by means of the thin-film approximation~\citep{ben2001fingering,joanny2012drop,Loisy2019PRL,trinschek2020thin,Loisy2020,shankar2022optimal,ioratim2022nonlinear,ford2024pattern}, which assumes that all variables describing the drop vary less in the direction parallel to the substrate than in the direction normal to it~\citep{oron1997long,craster2009dynamics}. However, this approximation is not accurate when the shape, the order of active units, and the self-generated flows vary arbitrarily in all directions, as it is the case in many living and synthetic systems. 

We revisit the paradigmatic continuum model of an active drop—containing a suspension of active units that exhibit nematic order and spontaneously self-generate flow—attached to a rigid, impermeable, flat surface. Unlike previous work, we do not make the thin-film approximation, but instead perform time-dependent simulations of the complete conservation equations. We find that the drop can exhibit a rich array of steady-state shapes and internal flows, which are a function of the alignment of active units with the boundaries and the active Capillary number comparing the active stresses exerted by the units and the capillary pressure. We uncover all possible states of the drop and the necessary conditions for symmetry breaking, which is determined by the boundary alignment of units only. These results are in stark contrast to those generated using the thin-film approximation, which predicts a unique equilibrium shape for the absolute value of the Capillary number, independent of alignment conditions and the contractile or extensile nature of the active stresses. Finally, we show that these equilibrium drop shapes can be controlled reversibly via surface anchoring and through the bulk active stresses exerted by the units.

%To unravel the interplay between interfacial morphodynamics, nematic order, and active flows, here we consider a minimal continuum model that describes an active drop anchored to a rigid, impermeable flat substrate, containing a suspension of active units that exhibit nematic order and spontaneously self-generate flow. This configuration is relevant to both living drop-like systems and synthetic drops that are attached to solid substrates, e.g., surface-attached microbial colonies~\citep{beroz2018verticalization,Pearce2019,hartmann2019emergence,nijjer2021mechanical,nijjer2023biofilms}, single-cell and tissue migration on surfaces~\citep{Tjhung2012,Tjhung2015,perez2019active}, and active wetting of synthetic and living systems~\citep{khalil2014active,alert2018role,perez2019active,aggarwal2023activity,Aggarwal2023PRL,liese2023chemically}.

Altogether, our findings provide a deeper understanding of the morphodynamics of active drops, with implications for various biological processes such as cell migration, tissue morphogenesis, biofilm development, and intracellular flows. These insights not only shed light on the complex interplay between order, flow, and shape in living systems but also offer a framework for the design and control of synthetic active materials and living systems.\\

\begin{figure*}[ht!]
  \begin{center}
   \includegraphics[width=\textwidth]{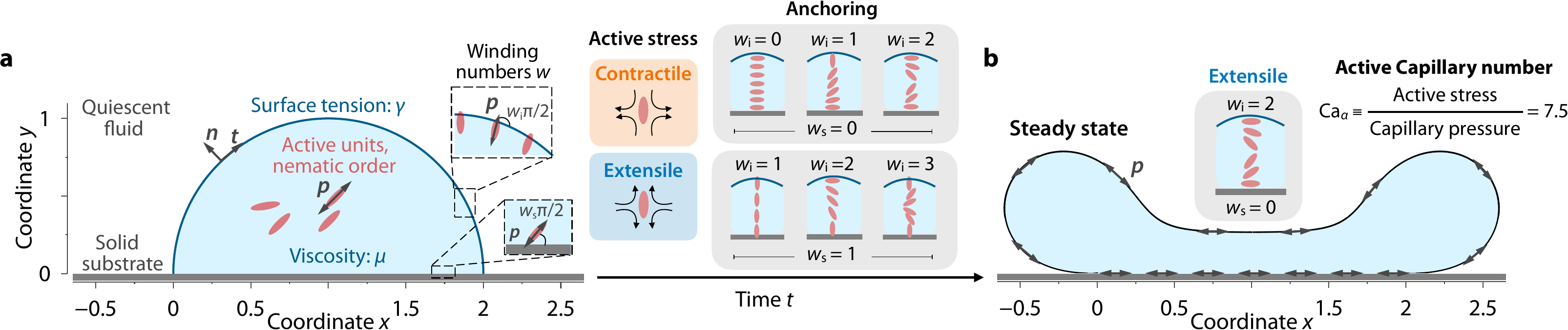}
    \caption{\label{fig:figure1} \textbf{Surface-attached active drops adopt stable steady-state shapes and internal flows}. \textbf{a}, Initial shape of a 2D liquid drop of viscosity $\mu$ and surface tension $\gamma$, attached to a rigid, impermeable substrate. The drop contains a uniform suspension of active units that exhibit nematic orientational order and self-generate flow through active stresses. These stresses can either be contractile, producing compressive flow along the axis of the units, or extensile, resulting in elongational flow. The vector field $\boldm{p}(\boldm{r},t)$ characterizes the orientational order of the units. These units can form an angle of $w_{\textrm{s}} \pi/2$ with the solid substrate and an angle of $w_{\textrm{i}} \pi/2$ with the liquid-air interface. Here, $w$ denotes the winding numbers, which represent the number of quarter turns the units make with respect to the direction tangential to the boundaries. The possible anchoring configurations include: $w_{\textrm{s}} = 0$ and $w_{\textrm{i}} = {0,1,2}$ for planar substrate anchoring, and $w_{\textrm{s}} = 1$ and $w_{\textrm{i}} = {1,2,3}$ for orthogonal (homeotropic) substrate anchoring. \textbf{b}, Stresses induced by activity spontaneously generate flow, causing deformation of the drop. Over time, the surface-attached active drop reaches a steady-state shape and flow, where active, viscous, and surface tension forces are in balance. This state depends on three dimensionless parameters: (i) the active Capillary number $\Ca_{\alpha}$, which compares the active stresses, which tend to deform the drop, with the capillary pressure, which tends to maintain a semicircular shape by minimizing the surface area per unit length; and (ii, iii) the orientation of the units, characterized by the winding numbers $w_{\textrm{s}}$ and $w_{\textrm{i}}$. Here, $\Ca_{\alpha} = 7.5$, $w_{\textrm{s}} = 0$, and $w_{\textrm{i}} = 2$, implying that the flow is extensile, and the units are tangentially aligned with both boundaries, making two quarter turns from the substrate to the liquid-air interface (Supplementary Movie 1).}
  \end{center}
\end{figure*}
%%%%%%%%%%%%%%%%%%%%%%%%%%%%%%%

%In particular, the ability of active matter to convert free energy into motion has been studied in detail in different living and synthetic systems~\citep{Marchetti2013,Tjhung2012,Tjhung2015}. \com{Complete with references of Tjhung \& Cates on crawling cells}

%The wetting dynamics of passive drops has been extensively studied~\citep{de1985wetting,bonn2009wetting,de2013capillarity}

%For simplicity, we consider that the flow generated by active stresses is incompressible and Newtonian, inertia is negligible, and the ambient fluid surrounding the drop is quiescent. In addition, the nematic units relax toward a energy minimum much faster than the flow time scale, and whose orientation only depends on the anchoring with the substrate and the deformable interface, but not by flow, which is usually referred to as the \textit{strong} elastic limit~\citep{de1993}.

\noindent \textbf{Model of an active nematic drop attached to a solid substrate}. We investigate the morphodynamics of surface-attached viscous nematic droplets: drops of fluid containing a suspension of elongated active units that acquire orientational order with head-tail symmetry. These units tend to maintain such order and exert force dipoles that spontaneously generate flow, thereby deforming the interface between the active fluid inside the drop and the surrounding fluid, which we consider to be passive and quiescent (Fig.~\ref{fig:figure1}a). For simplicity, we consider a two-dimensional (2D) continuum model to describe a nematic droplet of radius $R$ of an incompressible, Newtonian fluid of dynamic viscosity $\mu$ and surface tension $\gamma$ associated with the liquid-air interface. We assume that the droplet is attached to a rigid, impermeable substrate, with an initial semicircular shape, and contains a uniform suspension of active units. The nematic orientational order of the units is described by the director vector field $\boldm{p}(\boldm{r},t)$, where $\boldm{r}$ is position and $t$ is time. The flow generated by the active stress exerted by the units is described by the fluid isotropic pressure $\Pi(\boldm{r},t)$ and the velocity field $\boldm{u}(\boldm{r},t)$, assuming Stokes flow, i.e., low-Reynolds-number flow. The dimensionless volume and momentum conservation equations read:
\begin{equation}\label{eq:continuity_momentum}
\bnabla \bcdot \boldm{u} = 0, \quad \text{and} \quad \boldm{0} = \bnabla \bcdot \boldm{\sigma},
\end{equation}
where $\boldm{\sigma}$ is the stress tensor of the fluid. For simplicity, we consider that active units do not proliferate. The stress tensor takes into account the fluid isotropic pressure $\Pi$, which enforces flow incompressibility (Eq.~\ref{eq:continuity_momentum}), the viscous stress, and the active stress generated by the active components in the drop. Assuming that the flow is Newtonian and neglecting any complex rheology of the fluid resulting from the presence of the active units, the stress tensor is given by:
%We consider that the active contribution to the stress is proportional to the orientation vector field of the particles $\boldm{p}$, as follows,
\begin{equation}\label{eq:stresstensor}
\boldm{\sigma} = -\Pi \mathbfsf{I} + \mu \left[ \bnabla \boldm{u} + (\bnabla \boldm{u})^{\textrm{T}}\right] - \alpha \boldm{p} \boldm{p},
\end{equation}
where $\alpha$ is the signature of the active stress, which quantifies the strength of the force dipoles exerted by the active units. The sign of $\alpha$ characterizes the nature of the flow produced by the active units: for $\alpha > 0$, the flow is \textit{extensile}, while for $\alpha < 0$, it is \textit{contractile}, favoring stretching or contraction along the axis of the active unit, respectively. In a biological context, swimming microorganisms can produce either contractile or extensile flow depending on their motility mode~\citep{Marchetti2013}. Similarly, microtubule suspensions can exhibit either extensile or contractile behavior~\citep{keber2014topology,needleman2017active,lee2021myosin,berezney2022extensile}.

%bacteria such as \textit{Escherichia coli} produce extensile flows, while the eukaryotic algae \textit{Chlamydomonas reinhardtii} produce contractile flows. Additionally, microtubule suspensions can exhibit either extensile or contractile behavior~\citep{keber2014topology,lee2021myosin,berezney2022extensile}.

%%%%%%%%%%% Figure2: Map %%%%%%%%%%%
\begin{figure*}[ht!]
  \begin{center}
    \includegraphics[width=\textwidth]{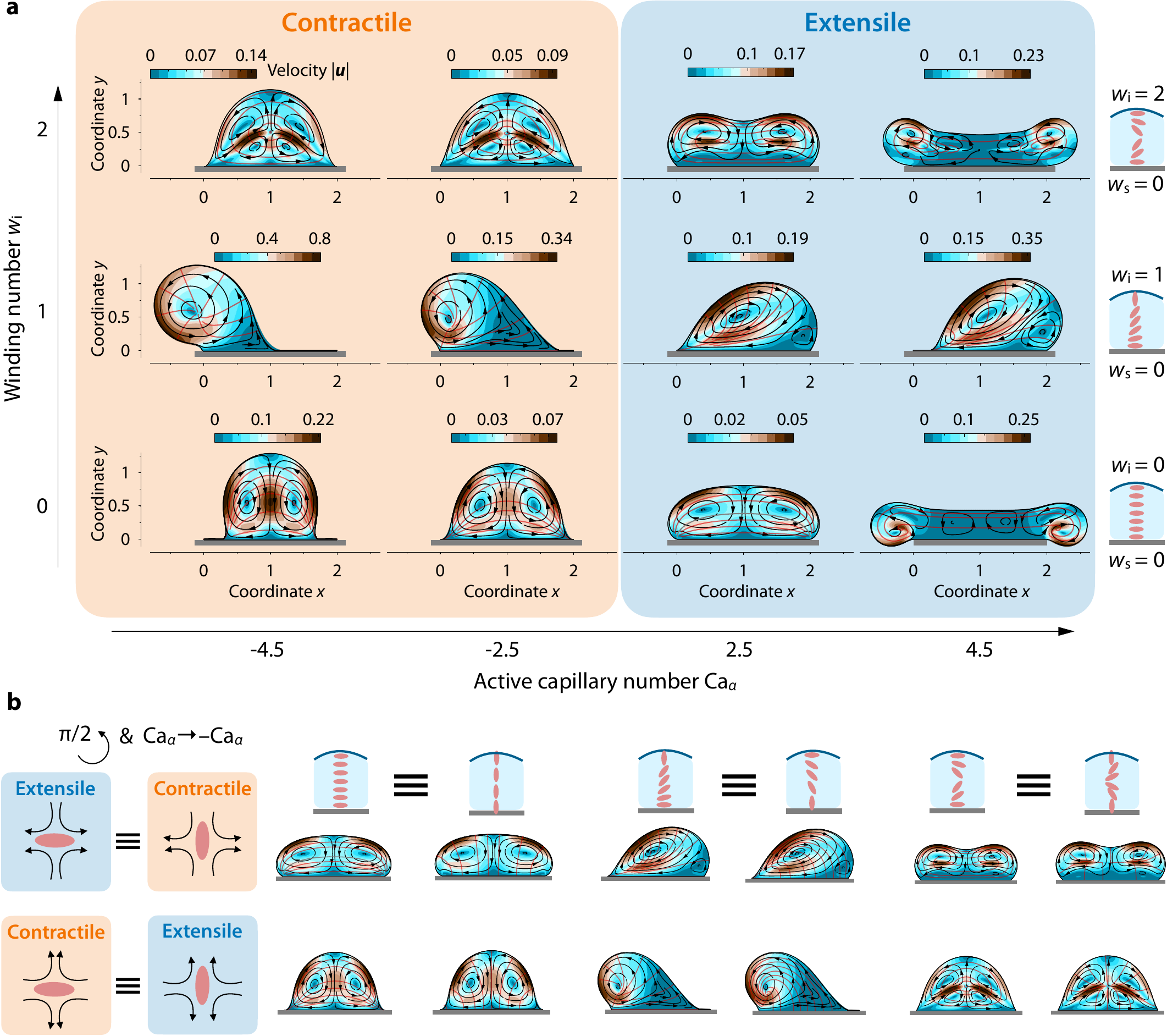}
    \caption{\label{fig:figure2} \textbf{Surface-attached active drops acquire a rich array of equilibrium shapes and flows}. \textbf{a}, Steady-state shapes and self-generated flows of a 2D active nematic drop attached to a rigid, planar substrate, as a function of the active capillary number $\Ca_{\alpha}$ and the interfacial winding number $w_{\textrm{i}}$, for planar substrate anchoring, $w_{\textrm{s}} = 0$ (Supplementary Movies 2-13). The color plots display the magnitude of the flow velocity $|\boldm{u}|$ inside the drop. The black curves represent the flow streamlines, and the grey curves depict the orientation of the active units. \textbf{b}, The steady states of a surface-attached active drop for planar ($w_{\textrm{s}} = 0$) and orthogonal ($w_{\textrm{s}} = 1$) unit anchoring with the substrate are equivalent under the transformations $w \to w + \pi/2$ and $\Ca_{\alpha} \to -\Ca_{\alpha}$. Here, $|\Ca_{\alpha}|=2.5$.}
  \end{center}
\end{figure*}
%%%%%%%%%%%%%%%%%%%%%%%%%%%%%%%

To determine the orientation field $\boldm{p}$ describing the nematic order of active units in the drop, we assume that this order is established much more quickly than the flows generated within the drop. This assumption implies that the orientation field relaxes instantaneously toward its free-energy minimum and is uncoupled from the flow dynamics, usually referred to as the strong elastic limit. We consider the following free energy for the orientation field~\citep{stephen1974,de1993,Prost2015}: 
\begin{equation}
    F = \int_A  \textrm{d} A \, \frac{K}{2} \left(\bnabla \boldm{p} : \bnabla \boldm{p}\right),
\end{equation}
where $K$ is the Frank elastic constant in the one-constant approximation~\citep{stephen1974,de1993} and $A$ is the area of the 2D drop. The first variation of $F$ with respect to $\boldm{p}$ yields the following equation for the instantaneous orientation field, 
\begin{equation}
\bnabla^2 \boldm{p} = \boldm{0}.
\end{equation}
At the interface of the drop we impose a kinematic condition specifying that the velocity of the interface is equal to the velocity of the fluid, thus precluding mass transfer across the interface, the stress balance, and the orientation of the units:
\begin{subequations}\label{eq:kinematic_surfacebalance}
\begin{align}
& \textrm{Fluid:} \nonumber & \\
& (\partial_t \boldm{r}_{\textrm{i}} - \boldm{u})\bcdot \hat{\boldm{n}} = 0, \, \, \boldm{\sigma} \bcdot \hat{\boldm{n}} = -\gamma \mathcal{C}\hat{\boldm{n}}
 \, \, \,  \text{at} \, \, \,  \boldm{r} = \boldm{r}_{\textrm{i}}, & \\
& \textrm{Active units:} \, \, 
\boldm{p} = \mathbfsf{R}_{\text{i}} \bcdot \hat{\boldm{t}} \, \,  \textrm{at} \, \, \boldm{r} = \boldm{r}_{\textrm{i}},
\end{align}
\end{subequations}
where $\boldm{r}_{\textrm{i}}$ is the position of the interface, $\hat{\boldm{n}}$ and $\hat{\boldm{t}}$ are the unit normal and tangential vectors to the interface, respectively, and $\mathcal{C} \equiv \bnabla \bcdot \hat{\boldm{n}}$ is twice the mean curvature of the liquid-air interface. Here, $\mathbfsf{R}_{\textrm{i}}$ is the rotation matrix prescribing the angle $\theta_{\textrm{i}}$ that the orientation of the units form with the tangent vector to the interface. For simplicity, we follow Refs.~\citep{Loisy2019PRL,Loisy2020} and consider only quarter turns in the orientation angle, i.e., $\theta_{\textrm{i}} = w_{\textrm{i}} \pi/2$, where $w \in \mathbb{Z}$ is the winding number (Fig.~\ref{fig:figure1}a).

%Regarding the boundary conditions for $\boldm{p}$, we consider that the orientation field is aligned with the substrate, and that forms a certain angle, $\theta$, with the interfacial tangent vector, $\boldm{t}$ (Fig.~\ref{fig:figure1}a)
%\begin{subequations}\label{eq:bc_p}
%\begin{gather}
%\boldm{p} = \boldm{e}_x \, \,  \text{at} \, \,  y = 0, \, \, \text{and}   
%\, \, \boldm{p} = \mathbfsf{R} \bcdot \boldm{t} \quad \textrm{at} \quad \boldm{r} = \boldm{r}_{\textrm{s}},
%\end{gather}
%\end{subequations}
%where $\mathbfsf{R}$ is the standard rotation matrix. In particular, following Refs.~\citep{Loisy2019PRL,Loisy2020}, here we only consider that the angle is $\theta = w \pi/2$ with $\omega \in \mathbb{Z}$ being the winding number, thus it can only rotate around $\boldm{t}$ counterclockwise in quarter turns. 

%The particular case of $w = 1$ is usually referred to as \textit{homeotropic condition}, which corresponds to a polarity field aligned with $\boldm{n}$ at the interface. (\com{discuss in terms of biological examples})

At the solid substrate, we impose no-permeation and no-slip conditions for the velocity field, i.e., $\boldm{u} = \boldm{0}$ at $y = 0$, and we prescribe the orientation of the active units with respect to the tangent vector to the substrate, equivalently to the condition at the interface, i.e., $\boldm{p} = \mathbfsf{R}_{\textrm{s}} \bcdot \hat{\boldm{e}}_x$, where $\mathbfsf{R}_{\textrm{s}}$ is the corresponding rotation matrix prescribing the angle $\theta_{\textrm{s}}$ at the substrate. For simplicity, we also assume that the orientation angle with the substrate only changes in quarter turns, and thus it is specified by an additional winding number $w_{\textrm{s}}$. As initial conditions, we consider the shape of the drop is semicircular and the fluid in the drop is at rest, i.e., $\boldm{u}(\boldm{r},t=0) = \boldm{0}$ and $\Pi(\boldm{r},t = 0) = \gamma/R$.\\

%In addition, we specify the orientation of the active units with respect to the interface and solid substrate, and fix them over time.

%In contrast to previous works that considered the thin-film approximation, in which all the variables that describe the flow, shape, and orientational order vary slowly along the direction of the substrate, we do not \textit{a priori} impose the slenderness assumption and and perform full numerical simulations of the conservation equations~\eqref{eq:continuity_momentum}-\eqref{eq:kinematic_surfacebalance}.\\

%This active constituents are able to convert the energy available in the drop into systematic motion, which macroscopically gives rise to an active stress in the fluid, thus generating flow and deforming the drop. 

%For simplicity, we assume that the drop is embedded in a passive ambient fluid, and that the initial shape of the 2D drop is a semicircle. 

%To describe the flow generated by the active units inside the drop, we impose mass and momentum conservation, and assume that inertia is negligible. 

\noindent \textbf{Dimensionless parameters governing the morphodynamics of a surface-attached active drop}. Before solving Eqs.~\eqref{eq:continuity_momentum}-\eqref{eq:kinematic_surfacebalance}, we non-dimensionalize these equations to reduce the number of parameters. To this end, we choose the initial drop radius $R$ as the characteristic length scale, viscocapillary time $\mu R/\gamma$ as the characteristic time scale, $\gamma/\mu$ as the corresponding characteristic velocity scale, and capillary pressure $\gamma/R$ as the characteristic pressure scale. All results and variables hereafter are nondimensionalized by these quantities. For simplicity, we retain the same notation for the dimensionless variables. Non-dimensionalization reveals only one governing dimensionless parameter, the \textit{active Capillary number}:
\begin{equation}\label{eq:capillary}
\Ca_{\alpha} \equiv \frac{\alpha}{\gamma/R}: \, \dfrac{\textrm{Active stress}}{\textrm{Capillary pressure}},
\end{equation}
which compares the active stress exerted by the active units with the capillary pressure~\citep{Giomi2014,Loisy2019PRL}. Thus, our minimal model is described by three dimensionless parameters: the active Capillary number, $\Ca_{\alpha}$, and the two winding numbers $w_{\textrm{i}}$ and $w_{\textrm{s}}$, which specify the orientation angles of the active units with respect to the liquid-air interface and solid substrate, respectively.

% This parameter is the active counterpart of the conventional Capillary number, which compares viscous stresses with capillary pressure

Previous works described surface-attached nematic drops using the thin-film approximation. This approximation simplifies Eqs.~\eqref{eq:continuity_momentum}-\eqref{eq:kinematic_surfacebalance} into a single equation describing the height of the drop $h(\boldm{r},t)$ by assuming that all variables change slowly along the direction of the substrate compared to the variation in the direction normal to the substrate: $\partial_t h + \bnabla \bcdot (h^3 \bnabla  \bnabla^2 h/3+\Ca_{\alpha,\text{lub}}h^2\boldm{r})$. Moreover, this approach reduces the number of dimensionless parameters to one, which is a modified active Capillary number that absorbs a winding number $w$ determining the orientational turns of the active units from the solid substrate to the liquid-air interface, $\Ca_{\alpha,\text{lub}} \equiv \alpha/(2 \pi w \gamma/R)$. In contrast, we do not \textit{a priori} impose the slenderness of the drop; instead, we conduct full numerical simulations of the conservation equations~\eqref{eq:continuity_momentum}-\eqref{eq:kinematic_surfacebalance}.\\

%As mentioned above, the thin-film approximation reduces the complexity of the system to a single equation describing the height of the drop $h(x,t)$: $\partial_t h + \partial_x (h^3 \partial_{xxx}h/3+\Ca_{\alpha,\text{lub}}h^2)$, where $x$ is the coordinate along the solid substrate. Furthermore, this approximation reduces the number of dimensionless parameters to one, which is a modified active Capillary number that absorbs a winding number $w$ determining the orientational turns of the active units from the solid substrate to the liquid-air interface, $\Ca_{\alpha,\text{lub}} \equiv \alpha/(2 \pi w \gamma/R)$.\\

%%%%%%%%%%% Figure3: velocities %%%%%%%%%%%
\begin{figure}[ht!]
  \begin{center}
  \includegraphics[width=0.48\textwidth]{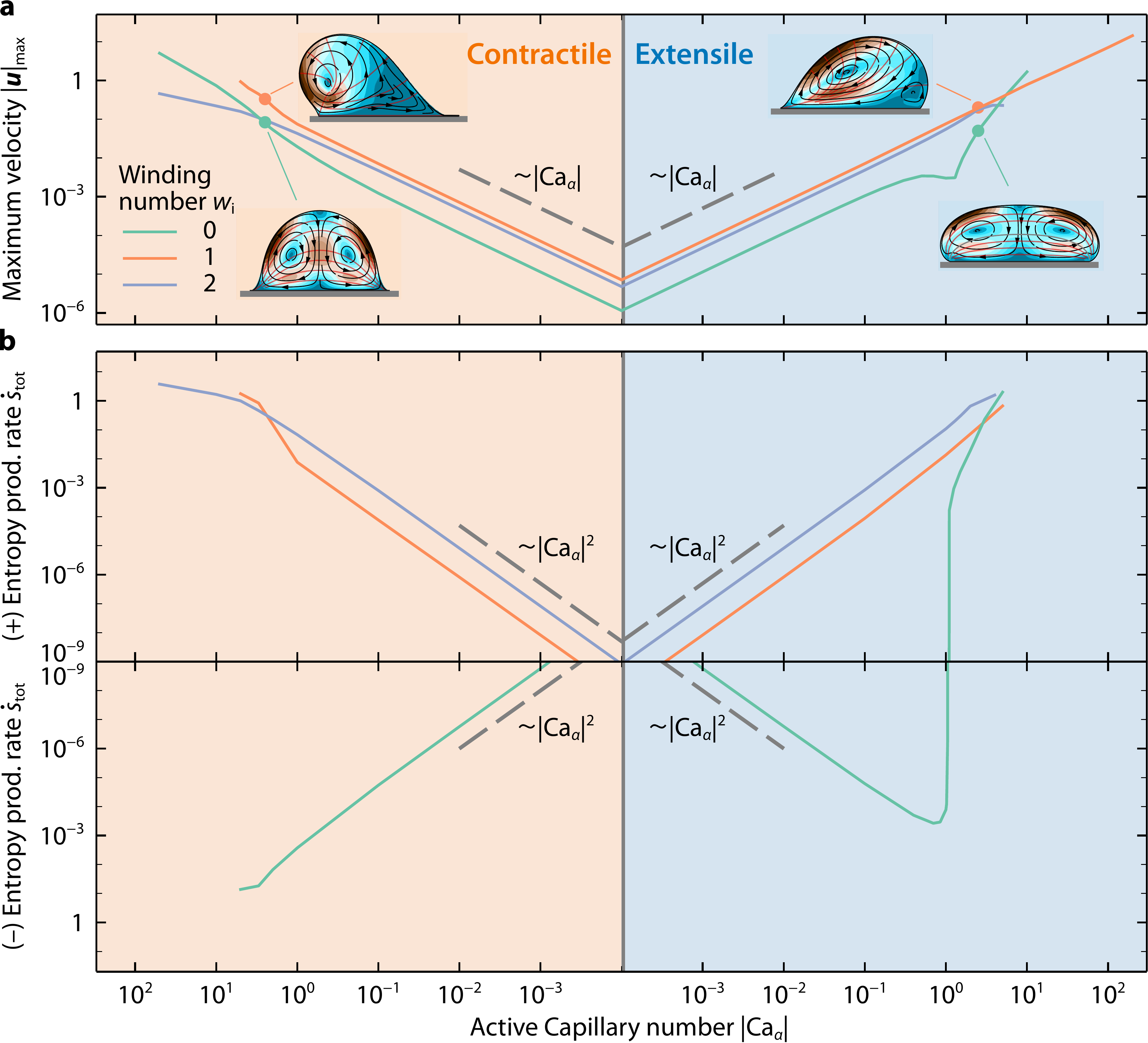}
      \caption{\textbf{Flow and dissipation inside surface-attached active drops follow simple scaling laws}. \textbf{a}, Maximum velocity $|\boldm{u}|_{\text{max}}$ inside the active drop as a function of the active Capillary number $\Ca_{\alpha}$ for contractile (left) and extensile (right) active stresses, and different values of the interfacial winding number $w_{\textrm{i}}$, and $w_{\textrm{s}} = 0$. \textbf{b}, Total entropy production rate $\dot{s}_{\text{tot}}$ of the active drop as a function of $\Ca_{\alpha}$, under the same conditions as in panel a. \label{fig:figure3}}
  \end{center}
\end{figure}
%%%%%%%%%%%%%%%%%%%%%%%%%%%%%%%

% Recent studies within the context bacterial biofilms have considered growth~\citep{Pearce2019,Qin2020}

%%%%%%%%%%%%%%%%% SECTION: Discussion %%%%%%%%%%%%%%%%%%%
\noindent \textbf{Surface-attached active drops reach stable steady states}. How is the organization of active nematic units influenced by, and in turn, how does it influence the interfacial morphodynamics of a surface-attached nematic drop? To address this question we perform numerical simulations of Eqs.~\eqref{eq:continuity_momentum}-\eqref{eq:kinematic_surfacebalance}. We first explore the case shown in Fig.~\ref{fig:figure1}b as an illustrative example, which corresponds to $\Ca_{\alpha} = 7.5$, $w_{\textrm{s}} = 0$, and $w_{\textrm{i}} = 2$ (Supplementary Movie 1). In this case, the active units generate extensile flows and their orientation rotates by $\pi$ radians from the substrate to the liquid-air interface. This spatial organization of the active units, combined with the high active stresses they exert, generates strong flows capable of inducing large deformations of the drop. However, after a transient period, the drop eventually reaches a stable, steady-state shape and internal flow, where active, viscous, and capillary forces are balanced, as shown in the snapshot in Fig.~\ref{fig:figure1}b. This equilibrium shape and flow contrasts with the thin-film prediction~\citep{Loisy2019PRL,Loisy2020}. Under this approximation, the active droplet always breaks symmetry, regardless of the value of the slender active Capillary number $\Ca_{\alpha,\text{lub}}$ or the orientation of the units relative to the substrate and liquid-air interface.

Motivated by this finding, we explore whether a surface-attached nematic drop always reaches a stable steady state for all values of the governing parameters, and investigate the necessary conditions for symmetry breaking, which can ultimately lead to drop migration.\\

%result is in stark contrast to previous works on active nematic drops in which a fingering instability is predicted~\citep{ben2001fingering,alert2022fingering}. 
 
%Under this approximation, the shape of the drop is only a function of the absolute value of the Capillary number $\Ca_{\alpha}$. Thus, the drop acquires the same shape for different values of the winding number $w$ and the extensile or contractile nature of the self-generated flow, if the absolute value of $\Ca_{\alpha}$ is fixed.\\ 

%In addition, the thin-film equation predicts that the drop \textit{always} breaks the symmetry, for all values of $\Ca_{\alpha}$.\\

%without developing an interfacial instability.\\

%%%%%%%%%%% Figure4: Control
\begin{figure*}[ht!]
  \begin{center}
 \includegraphics[width=0.95\textwidth]{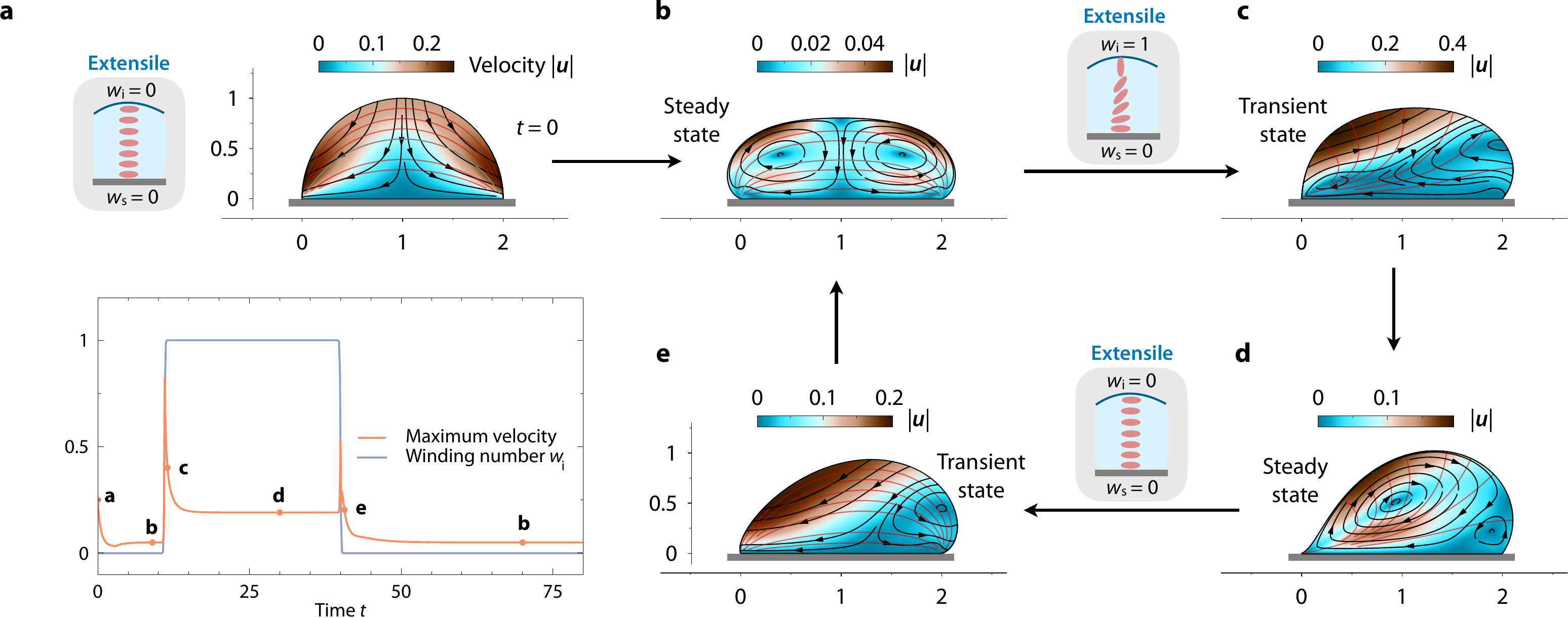}
    \caption{\textbf{The morphodynamics of surface-attached active drops can be controlled reversibly through the anchoring of active units}. \textbf{a}, Initial shape and flow of the drop with $w_{\textrm{s}} = 0$, $w_{\textrm{i}} = 0$, and $\Ca_{\alpha} = 2.5$ (top). Interfacial winding number $w_{\textrm{i}}$ and maximum velocity $|\boldm{u}|_{\text{max}}$ as functions of time (bottom). The dots indicate the times shown in panels b-e. \textbf{b}, Steady state of the drop with $w_{\textrm{s}} = 0$, $w_{\textrm{i}} = 0$. \textbf{c}, Transient state of the drop immediately after $w_{\textrm{i}}$ is changed drastically from $w_{\textrm{i}} = 0$ to $w_{\textrm{i}} = 1$, at which point the drop begins to break symmetry. \textbf{d}, After a transient period, the drop reaches steady state with $w_{\textrm{i}} = 1$. \textbf{e}, Transient state of the drop after $w_{\textrm{i}}$ is changed back to $w_{\textrm{i}} = 0$. After another transient period, the drop achieves the same steady state shown in panel b. \label{fig:figure4}}
  \end{center}
\end{figure*}
%%%%%%%%%%%%%%%%%%%%%%%%%%%%%%%

\noindent \textbf{Surface-attached active drops adopt a rich array of equilibrium shapes and flows}. What are the possible shapes that a surface-attached active drop can adopt? To address this question, we perform time-dependent simulations for all possible combinations of winding numbers $w_{\textrm{i}}$ and $w_{\textrm{s}}$, extensile and contractile stresses, and values of the active Capillary number $|\Ca_{\alpha}|$. The possible anchoring configurations are $w_{\textrm{s}} = 0$ and $w_{\textrm{i}} = \{0,1,2\}$ for planar substrate anchoring, and $w_{\textrm{s}} = 1$ and $w_{\textrm{i}} = \{1,2,3\}$ for homeotropic substrate anchoring, as shown in the schematic in Fig.~\ref{fig:figure1}. We find that the drop adopts equilibrium shapes for all values of $\Ca_{\alpha}$ and $w$ (Supplementary Movies 2-13). This result contrasts with previous works on active nematic drops that are not attached to a solid boundary, which have shown that such free drops undergo fingering instabilities for different anchoring conditions and values of $\Ca_{\alpha}$~\citep{ben2001fingering,alert2022fingering} (Supplementary Movies 14 and 15). Moreover, the drop exhibits a wide variety of stable steady states, illustrated in the  $(\Ca_{\alpha},w)$ state diagram in Fig.~\ref{fig:figure2}. Among these, some states spontaneously break symmetry. What conditions are necessary for this symmetry breaking?

\textit{Parallel substrate anchoring}. We begin by analyzing cases where the units align parallel to the solid substrate, corresponding to $w_{\textrm{s}} = 0$ (Fig.~\ref{fig:figure2}a). Additionally, we consider the case of $w_{\textrm{i}} = 0$, where the units are also aligned parallel to the liquid-air interface (Fig.~\ref{fig:figure2}a, bottom row; Supplementary Movies 2-5). These anchoring conditions generate a bipolar structure in the ordering of units, with two point defects at the triple contact points between the solid substrate, the fluid, and the surrounding environment. Consequently, the order of active units, the flow, and the shape of the drop are symmetric with respect to its central midplane. Interestingly, for $-4.5 \lesssim \Ca_{\alpha} \lesssim 2.5$, this nematic ordering produces two steady-state counter-rotating vortices, whose handedness is independent of the extensile or contractile nature of the active stresses. However, the drop shape is markedly different between the contractile ($\Ca_{\alpha} < 0$) and extensile ($\Ca_{\alpha} > 0$) cases. In the former, the drop contracts in the direction parallel to the substrate, resulting in a \textit{mushroom-like} shape. In the latter, the extensile stresses tend to flatten the drop in the $x$. Intriguingly, for $\Ca_{\alpha} \gtrsim 4.5$, the extensile stresses become strong enough to elongate and shrink the drop at its center, generating a flat film that connects two lobes with multiple counter-rotating vortices. This shape transition coincides with the disappearance of the point defects at the triple contact point and the emergence of a spiral defect inside the two lobes. These results contrast with the thin-film prediction, in which the condition $w_{\textrm{s}} = w_{\textrm{i}} = 0$ does not generate flow and the drops remain undeformed. This is a consequence of neglecting the variation of $\boldm{p}$ along the direction of the substrate.

%This anchoring conditions do not break the spatial symmetry of nematic order, and thus both the shape of the drop and the internal flow also remain symmetric. 

We next focus on the case with $w_{\textrm{s}} = 0$ and $w_{\textrm{i}} = 1$, in which the units make a counterclockwise quarter turn from the substrate to the liquid-air interface (Fig.~\ref{fig:figure2}a, middle row; Supplementary Movies 6-9). Our results reveal that a mismatch in the orientation of the units between these boundaries is a necessary condition for symmetry breaking. The direction in which the drop breaks symmetry depends on the handedness of the units' rotation and the extensile and contractile nature of the active stresses. If the units make a clockwise rotation ($w_{\textrm{s}} = 0$ and $w_{\textrm{i}} = 3$), the shape and flow of the drop are identical to those in the case of counterclockwise rotation in Fig.~\ref{fig:figure2}a under a $x \to -x$ mirror symmetry. Furthermore, for the same units' rotation, the extensile and contractile cases lead to opposite flow vortex rotations. As a consequence, symmetry is broken in opposite directions. Intriguingly, the hedgehog defect in the units' ordering at one of the poles of the drop, observed in the extensile case, does not emerge in the contractile case, where instead a stable spiral defect appears in the bulk of the drop. Again, the asymmetric steady-state shapes and flows obtained here are markedly different from the thin-film prediction. Under this approximation, the extensile and contractile cases are identical under a $x \to - x$ reflection, and changing the handedness of the units' rotation is equivalent to changing the sign of $\Ca_{\alpha,\textrm{lub}}$, which does not hold in the complete model.

Next, we explore the case of $w_{\textrm{s}} = 0$ and $w_{\textrm{i}} = 2$ (Fig.~\ref{fig:figure2}, top row; Supplementary Movies 10-13), where the units rotate $\pi$ radians from the substrate, where they are horizontally aligned, to the interface, where they are also tangentially aligned. As expected, this configuration does not break the symmetry of the drop, unlike the thin-film equation. In particular, these anchoring conditions generate a vortex defect in the ordering of active units. In contractile cases, this defect generates a flow with two pairs of counter-rotating vortices, and the drop achieves a mushroom-like shape, similar to the case with $w_{\textrm{i}} = w_{\textrm{s}} = 0$. In extensile cases, the drop becomes more elongated, and high activity turns the vortex defect into a line defect, resulting in a complex flow structure with multiple vortices, similar to the case with $w_{\textrm{i}} = w_{\textrm{s}} = 0$.

\textit{Homeotropic substrate anchoring}. Thus far, we have explored cases in which the active units are aligned tangentially with the solid substrate ($w_{\textrm{s}} = 0$). To investigate whether other substrate anchoring configurations can alter the drop's morphodynamics, we conduct numerical simulations of drops in which the units are oriented orthogonally to the substrate ($w_{\textrm{s}} = 1$). We find that the drop shapes obtained with $(w_{\textrm{s}},w_{\textrm{i}}) = \{(0,0),(0,1),(0,2)\}$ are equivalent to those obtained by a $\pi/2$ rotation of the units, i.e., $(w_{\textrm{s}},w_{\textrm{i}}) = \{(1,1),(1,2),(1,3)\}$, under the transformation $\Ca_{\alpha} \to -\Ca_{\alpha}$. This result makes sense intuitively, as the flow generated by an extensile unit oriented horizontally is equivalent to the one generated by a contractile unit oriented vertically, and vice versa (Fig.~\ref{fig:figure1}b). This finding is unlikely to hold if the strong elastic limit is relaxed, i.e., if the flow is coupled back with the orientation of the active units. Exploring the influence of this additional complexity will be a useful direction for future work.\\

\noindent \textbf{Flow and energy dissipation in active drops follow simple scaling laws}. Despite the complexity of shapes and internal flows exhibited by a surface-attached active nematic drop, steady-state flow inside the drop follows simple scaling laws (Fig.~\ref{fig:figure3}). In particular, the maximum velocity, $\text{max}(|\boldm{u}|) \equiv |\boldm{u}|_{\text{max}}$, follows identical linear scaling laws for both contractile and extensile cases when $|\Ca_{\alpha}| \lesssim 1$, i.e., $|\boldm{u}|_{\text{max}} \sim |\Ca_{\alpha}|$, for any value of $w_{\text{i}}$. Interestingly, the maximum velocity is usually higher when the drop breaks symmetry ($w_{\text{i}} = 1$), except in certain cases at large values of $|\Ca_{\alpha}|$, where the drop undergoes large deformations (e.g., $w_{\text{i}} = 0$ and $\Ca_{\alpha} \gtrsim 5 $).

What is the energy expenditure of the active drop? We compute the total entropy production rate as follows:
\begin{equation}\label{eq:entropy}
    \rho T \dot{s}_{\text{tot}} = \rho T \int_A \text{d} A \dot{s} =  \int_{A} \text{d}A \boldm{\sigma}\textbf{:}\bnabla \boldm{u},
\end{equation}
where $\rho$ is the fluid density, $T$ is the temperature, and $\dot{s} = \partial_t s+\boldm{u}\bcdot \bnabla s$. The total entropy production can decomposed into two contributions, i.e., the dissipation associated with viscous stresses, and the work done by the active units:
\begin{equation}\label{eq:entropy_contrib}
    \dot{s}_{\text{v}} = \int_{A} \text{d}A \boldm{\sigma}_{\text{v}} \textbf{:} \bnabla \boldm{u}, \quad \dot{s}_{\text{a}} = \int_{A} \text{d}A \boldm{\sigma}_{\text{a}} \textbf{:} \bnabla \boldm{u},
\end{equation}
where $\boldm{\sigma}_{\text{v}} = \mu[\bnabla \boldm{u}+\left(\bnabla \boldm{u}\right)^{\text{T}}]$ and $\boldm{\sigma}_{\text{a}} = -\alpha \boldm{p}\boldm{p}$ are the viscous and active components of the total stress tensor, respectively. We compute Eqs.~\eqref{eq:entropy}-\eqref{eq:entropy_contrib} for the case where units are tangentially aligned with the substrate, $w_{\text{s}} = 0$. Viscous dissipation is always irreversible, increasing the system's entropy, $\dot{s}_{\text{v}} > 0$. In our system, we find that the dimensionless viscous dissipation scales as $\dot{s}_{\text{v}} \sim \Ca_{\alpha}^2$ for all values of $w_{\text{i}}$ when $|\Ca_{\alpha}| \lesssim 1$, and is maximum when symmetry is broken, i.e., $w_{\text{i}} = 1$. Conversely, the active contribution scales as $\dot{s}_{\text{a}} \sim -\Ca_{\alpha}^2$ for all values of $w_{\text{i}}$ when $|\Ca_{\alpha}| \lesssim 1$. Interestingly, only when $w_{\text{i}} = 0$ does viscous dissipation remain smaller than the work done by active forces to sustain the self-generated flow and ordering, implying a negative value of $\dot{s}_{\text{tot}}$ (Fig.~\ref{fig:figure3}b, bottom). This behavior reverses only in the extensile case, $\Ca_{\alpha} > 0$, when the drop undergoes large deformations, i.e., when $\Ca_{\alpha} \gtrsim 1$, at which point $\dot{s}_{\text{tot}}$ becomes positive (Fig.~\ref{fig:figure3}b, top right).\\

%Although lubrication theory cannot accurately predict the steady states, this scaling law is in fair agreement with the thin-film prediction, $V \sim \Ca_{\alpha}/(2 \pi w_{\textrm{i}})$ for $w_{\textrm{i}} \neq 0$ and $w_{\textrm{s}} = 0$, which also shows that the velocity is maximum when the symmetry of the drop is broken.\\

%The fluid velocity in an active nematic drop follows simple scaling laws and is maximum when symmetry is broken.\\

%Can we reversibly control the steady-state shapes and flows shown in Fig.~\ref{fig:figure2} in our minimal model?

\noindent \textbf{Reversible shape and flow control through time-dependent anchoring of active units}. Our results thus far highlight the pivotal importance of interfacial anchoring on drop morphodynamics---suggesting a tantalizing route toward reversibly \textit{controlling} drop shape and flows. To explore this possibility, we vary the anchoring of the active units at the liquid-air interface over time (Fig.~\ref{fig:figure4}a, bottom). The upper panel of Fig.~\ref{fig:figure4}a shows the initial condition of the drop when the units are tangentially aligned with both the substrate and the liquid-air interface ($w_{\textrm{s}} = w_{\textrm{i}} = 0$), and $\Ca_{\alpha} = 2.5$. The anchoring with the substrate remains fixed, while the winding number at the liquid-air interface varies over time, as specified in the bottom panel of Fig.~\ref{fig:figure4}b, which also shows the maximum velocity inside the drop, $|\boldm{u}|_{\text{max}}$, to determine when the drop reaches steady state. At time $t = 10$, the drop reaches a stable steady-state shape and flow (Fig.~\ref{fig:figure4}b). At $t = 11$, we drastically change the interfacial winding number $w_{\textrm{i}}$ from $w_{\textrm{i}} = 0$ to $w_{\textrm{i}} = 1$, forcing the units to orient perpendicular to the liquid-air interface. After a transient period ($11\lesssim t \lesssim 15$; Fig.~\ref{fig:figure4}c), the drop acquires a new steady state (Fig.~\ref{fig:figure4}d). To test if this process is reversible, at $t = 40$, we drastically change the winding number back to $w_{\textrm{i}} = 0$. Although the transient states corresponding to $w_{\textrm{i}} = 1 \to w_{\textrm{i}} = 0$ (Fig.~\ref{fig:figure1}e) are not identical to its opposite, $w_{\textrm{i}} = 0 \to w_{\textrm{i}} = 1$ (Fig.~\ref{fig:figure1}c), implying that changing the boundary condition is not equivalent to time reversal, the system relaxes back to the same steady state over time (Fig.~\ref{fig:figure1}b). This result shows that, despite having a deformable liquid-air interface, this minimal active matter system allows for reversible control of stable steady-state shapes.\\

\noindent \textbf{Discussion}. Controlling the shape, flow, and interfacial morphodynamics of active matter is key to understanding the functioning of biological systems and designing functional soft active materials. Here, by considering a minimal model of a surface-attached active drop, we have shown that such an active system can exhibit a rich array of steady-state shapes and internal flows. Additionally, we showed that these states can be controlled reversibly, either through surface anchoring of the active components or by varying their activity strength. These features cannot be captured by the thin-film approximation, which is often used to describe such systems.

Despite the rich behaviors found in this active system, we made several simplifications and assumptions in formulating our theoretical description. For example, our model neglects inertial effects, advection of the active components by the self-generated flows, proliferation, and complex fluid rheology---features that are usually present in both synthetic and living systems. Similarly, here we neglected density variations of active components as well as complex chemo-mechanical interactions with the rigid substrate, the latter of which is known to play a crucial role in the wetting of biomolecular condensates. Exploring the impact of these additional complexities on the morphodynamics of active systems is a promising avenue for future research. Altogether, by deepening our understanding of the morphodynamics of active drops, we hope our results will inspire and inform new experiments in active systems.\\

%Active matter theories have been proven to be a powerful approach to understand the emergent collective phenomena and dynamics of living systems and synthetic active materials. Moreover. In this paper, we show that a minimal theoretical model of nematic droplet anchored to a solid substrate give rise to a rich array of shapes and internal flows that cannot be captured by the thin-film approximation.\\

%\begin{itemize}
%    \item Inertia and advection of units
%    \item Rheology
%    \item Proliferation (bacteria, mammalian cells, replicating systems)
%    \item Unbounded films

%    \item Relative importance of the advection of $\boldm{p}$ on the morphology of the drop
%\end{itemize}

%------------------------------------ACKNOWLEDGEMENTS-------------------------------------%

\noindent\textbf{Acknowledgements}. We thank the members of the Datta Lab for their valuable feedback. A.M.-C. acknowledges support from the Princeton Center for Theoretical Science, the Center for the Physics of Biological Function, and the Human Frontier Science Program through the grant LT000035/2021-C. S.S.D. acknowledges support from NSF Grants CBET-1941716 and DMR-2011750, as well as the Camille Dreyfus Teacher-Scholar and Pew Biomedical Scholars Programs.\\

\noindent \textbf{Data availability}. The data supporting the findings of this study are available from the corresponding authors upon request.\\

\noindent \textbf{Code availability}. The computer code used for simulations is available from the corresponding authors upon reasonable request.\\

\noindent \textbf{Competing interests}. The authors do not have any competing interests.\\

\noindent \textbf{Author contributions}. A.M.-C.: Conceptualization, Methodology, Software, Validation, Formal analysis, Investigation, Data Curation, Writing - Original Draft, Writing -Review \& Editing, Visualization, Project administration, Funding acquisition. S.S.D.: Conceptualization, Resources, Writing - Review \& Editing, Supervision, Project administration, Funding acquisition.

%--------------------------------------BIBLIOGRAPHY---------------------------------------%
%\bibliographystyle{}
\bibliography{biblio}

\begin{thebibliography}{48}%
\makeatletter
\providecommand \@ifxundefined [1]{%
 \@ifx{#1\undefined}
}%
\providecommand \@ifnum [1]{%
 \ifnum #1\expandafter \@firstoftwo
 \else \expandafter \@secondoftwo
 \fi
}%
\providecommand \@ifx [1]{%
 \ifx #1\expandafter \@firstoftwo
 \else \expandafter \@secondoftwo
 \fi
}%
\providecommand \natexlab [1]{#1}%
\providecommand \enquote  [1]{``#1''}%
\providecommand \bibnamefont  [1]{#1}%
\providecommand \bibfnamefont [1]{#1}%
\providecommand \citenamefont [1]{#1}%
\providecommand \href@noop [0]{\@secondoftwo}%
\providecommand \href [0]{\begingroup \@sanitize@url \@href}%
\providecommand \@href[1]{\@@startlink{#1}\@@href}%
\providecommand \@@href[1]{\endgroup#1\@@endlink}%
\providecommand \@sanitize@url [0]{\catcode `\\12\catcode `\$12\catcode `\&12\catcode `\#12\catcode `\^12\catcode `\_12\catcode `\%12\relax}%
\providecommand \@@startlink[1]{}%
\providecommand \@@endlink[0]{}%
\providecommand \url  [0]{\begingroup\@sanitize@url \@url }%
\providecommand \@url [1]{\endgroup\@href {#1}{\urlprefix }}%
\providecommand \urlprefix  [0]{URL }%
\providecommand \Eprint [0]{\href }%
\providecommand \doibase [0]{http://dx.doi.org/}%
\providecommand \selectlanguage [0]{\@gobble}%
\providecommand \bibinfo  [0]{\@secondoftwo}%
\providecommand \bibfield  [0]{\@secondoftwo}%
\providecommand \translation [1]{[#1]}%
\providecommand \BibitemOpen [0]{}%
\providecommand \bibitemStop [0]{}%
\providecommand \bibitemNoStop [0]{.\EOS\space}%
\providecommand \EOS [0]{\spacefactor3000\relax}%
\providecommand \BibitemShut  [1]{\csname bibitem#1\endcsname}%
\let\auto@bib@innerbib\@empty
%</preamble>
\bibitem [{\citenamefont {Marchetti}\ \emph {et~al.}(2013)\citenamefont {Marchetti}, \citenamefont {Joanny}, \citenamefont {Ramaswamy}, \citenamefont {Liverpool}, \citenamefont {Prost}, \citenamefont {Rao},\ and\ \citenamefont {Simha}}]{Marchetti2013}%
  \BibitemOpen
  \bibfield  {author} {\bibinfo {author} {\bibfnamefont {M.~C.}\ \bibnamefont {Marchetti}}, \bibinfo {author} {\bibfnamefont {J.-F.}\ \bibnamefont {Joanny}}, \bibinfo {author} {\bibfnamefont {S.}~\bibnamefont {Ramaswamy}}, \bibinfo {author} {\bibfnamefont {T.~B.}\ \bibnamefont {Liverpool}}, \bibinfo {author} {\bibfnamefont {J.}~\bibnamefont {Prost}}, \bibinfo {author} {\bibfnamefont {M.}~\bibnamefont {Rao}}, \ and\ \bibinfo {author} {\bibfnamefont {R.~A.}\ \bibnamefont {Simha}},\ }\href@noop {} {\bibfield  {journal} {\bibinfo  {journal} {Rev. Mod. Phys.}\ }\textbf {\bibinfo {volume} {85}},\ \bibinfo {pages} {1143} (\bibinfo {year} {2013})}\BibitemShut {NoStop}%
\bibitem [{\citenamefont {Zhang}\ \emph {et~al.}(2021)\citenamefont {Zhang}, \citenamefont {Mozaffari},\ and\ \citenamefont {de~Pablo}}]{zhang2021autonomous}%
  \BibitemOpen
  \bibfield  {author} {\bibinfo {author} {\bibfnamefont {R.}~\bibnamefont {Zhang}}, \bibinfo {author} {\bibfnamefont {A.}~\bibnamefont {Mozaffari}}, \ and\ \bibinfo {author} {\bibfnamefont {J.}~\bibnamefont {de~Pablo}},\ }\href@noop {} {\bibfield  {journal} {\bibinfo  {journal} {Nat. Rev. Mater.}\ }\textbf {\bibinfo {volume} {6}},\ \bibinfo {pages} {437} (\bibinfo {year} {2021})}\BibitemShut {NoStop}%
\bibitem [{\citenamefont {Hallatschek}\ \emph {et~al.}(2023)\citenamefont {Hallatschek}, \citenamefont {Datta}, \citenamefont {Drescher}, \citenamefont {Dunkel}, \citenamefont {Elgeti}, \citenamefont {Waclaw},\ and\ \citenamefont {Wingreen}}]{hallatschek2023proliferating}%
  \BibitemOpen
  \bibfield  {author} {\bibinfo {author} {\bibfnamefont {O.}~\bibnamefont {Hallatschek}}, \bibinfo {author} {\bibfnamefont {S.~S.}\ \bibnamefont {Datta}}, \bibinfo {author} {\bibfnamefont {K.}~\bibnamefont {Drescher}}, \bibinfo {author} {\bibfnamefont {J.}~\bibnamefont {Dunkel}}, \bibinfo {author} {\bibfnamefont {J.}~\bibnamefont {Elgeti}}, \bibinfo {author} {\bibfnamefont {B.}~\bibnamefont {Waclaw}}, \ and\ \bibinfo {author} {\bibfnamefont {N.~S.}\ \bibnamefont {Wingreen}},\ }\href@noop {} {\bibfield  {journal} {\bibinfo  {journal} {Nat. Rev. Phys.}\ ,\ \bibinfo {pages} {1}} (\bibinfo {year} {2023})}\BibitemShut {NoStop}%
\bibitem [{\citenamefont {Fern{\'a}ndez}\ \emph {et~al.}(2021)\citenamefont {Fern{\'a}ndez}, \citenamefont {Buchmann}, \citenamefont {Goychuk}, \citenamefont {Engelbrecht}, \citenamefont {Raich}, \citenamefont {Scheel}, \citenamefont {Frey},\ and\ \citenamefont {Bausch}}]{fernandez2021surface}%
  \BibitemOpen
  \bibfield  {author} {\bibinfo {author} {\bibfnamefont {P.}~\bibnamefont {Fern{\'a}ndez}}, \bibinfo {author} {\bibfnamefont {B.}~\bibnamefont {Buchmann}}, \bibinfo {author} {\bibfnamefont {A.}~\bibnamefont {Goychuk}}, \bibinfo {author} {\bibfnamefont {L.}~\bibnamefont {Engelbrecht}}, \bibinfo {author} {\bibfnamefont {M.}~\bibnamefont {Raich}}, \bibinfo {author} {\bibfnamefont {C.}~\bibnamefont {Scheel}}, \bibinfo {author} {\bibfnamefont {E.}~\bibnamefont {Frey}}, \ and\ \bibinfo {author} {\bibfnamefont {A.}~\bibnamefont {Bausch}},\ }\href@noop {} {\bibfield  {journal} {\bibinfo  {journal} {Nat. Phys.}\ }\textbf {\bibinfo {volume} {17}},\ \bibinfo {pages} {1130} (\bibinfo {year} {2021})}\BibitemShut {NoStop}%
\bibitem [{\citenamefont {Ishihara}\ \emph {et~al.}(2023)\citenamefont {Ishihara}, \citenamefont {Mukherjee}, \citenamefont {Gromberg}, \citenamefont {Brugu{\'e}s}, \citenamefont {Tanaka},\ and\ \citenamefont {J{\"u}licher}}]{ishihara2023topological}%
  \BibitemOpen
  \bibfield  {author} {\bibinfo {author} {\bibfnamefont {K.}~\bibnamefont {Ishihara}}, \bibinfo {author} {\bibfnamefont {A.}~\bibnamefont {Mukherjee}}, \bibinfo {author} {\bibfnamefont {E.}~\bibnamefont {Gromberg}}, \bibinfo {author} {\bibfnamefont {J.}~\bibnamefont {Brugu{\'e}s}}, \bibinfo {author} {\bibfnamefont {E.}~\bibnamefont {Tanaka}}, \ and\ \bibinfo {author} {\bibfnamefont {F.}~\bibnamefont {J{\"u}licher}},\ }\href@noop {} {\bibfield  {journal} {\bibinfo  {journal} {Nat. Phys.}\ }\textbf {\bibinfo {volume} {19}},\ \bibinfo {pages} {177} (\bibinfo {year} {2023})}\BibitemShut {NoStop}%
\bibitem [{\citenamefont {Beroz}\ \emph {et~al.}(2018)\citenamefont {Beroz}, \citenamefont {Yan}, \citenamefont {Meir}, \citenamefont {Sabass}, \citenamefont {Stone}, \citenamefont {Bassler},\ and\ \citenamefont {Wingreen}}]{beroz2018verticalization}%
  \BibitemOpen
  \bibfield  {author} {\bibinfo {author} {\bibfnamefont {F.}~\bibnamefont {Beroz}}, \bibinfo {author} {\bibfnamefont {J.}~\bibnamefont {Yan}}, \bibinfo {author} {\bibfnamefont {Y.}~\bibnamefont {Meir}}, \bibinfo {author} {\bibfnamefont {B.}~\bibnamefont {Sabass}}, \bibinfo {author} {\bibfnamefont {H.}~\bibnamefont {Stone}}, \bibinfo {author} {\bibfnamefont {B.}~\bibnamefont {Bassler}}, \ and\ \bibinfo {author} {\bibfnamefont {N.}~\bibnamefont {Wingreen}},\ }\href@noop {} {\bibfield  {journal} {\bibinfo  {journal} {Nat. Phys.}\ }\textbf {\bibinfo {volume} {14}},\ \bibinfo {pages} {954} (\bibinfo {year} {2018})}\BibitemShut {NoStop}%
\bibitem [{\citenamefont {Pearce}\ \emph {et~al.}(2019)\citenamefont {Pearce}, \citenamefont {Song}, \citenamefont {Skinner}, \citenamefont {Mok}, \citenamefont {Hartmann}, \citenamefont {Singh}, \citenamefont {Jeckel}, \citenamefont {Oishi}, \citenamefont {Drescher},\ and\ \citenamefont {Dunkel}}]{Pearce2019}%
  \BibitemOpen
  \bibfield  {author} {\bibinfo {author} {\bibfnamefont {P.}~\bibnamefont {Pearce}}, \bibinfo {author} {\bibfnamefont {B.}~\bibnamefont {Song}}, \bibinfo {author} {\bibfnamefont {D.~J.}\ \bibnamefont {Skinner}}, \bibinfo {author} {\bibfnamefont {R.}~\bibnamefont {Mok}}, \bibinfo {author} {\bibfnamefont {R.}~\bibnamefont {Hartmann}}, \bibinfo {author} {\bibfnamefont {P.~K.}\ \bibnamefont {Singh}}, \bibinfo {author} {\bibfnamefont {H.}~\bibnamefont {Jeckel}}, \bibinfo {author} {\bibfnamefont {J.~S.}\ \bibnamefont {Oishi}}, \bibinfo {author} {\bibfnamefont {K.}~\bibnamefont {Drescher}}, \ and\ \bibinfo {author} {\bibfnamefont {J.}~\bibnamefont {Dunkel}},\ }\href@noop {} {\bibfield  {journal} {\bibinfo  {journal} {Phys. Rev. Lett.}\ }\textbf {\bibinfo {volume} {123}},\ \bibinfo {pages} {258101} (\bibinfo {year} {2019})}\BibitemShut {NoStop}%
\bibitem [{\citenamefont {Hartmann}\ \emph {et~al.}(2019)\citenamefont {Hartmann}, \citenamefont {Singh}, \citenamefont {Pearce}, \citenamefont {Mok}, \citenamefont {Song}, \citenamefont {D{\'\i}az-Pascual}, \citenamefont {Dunkel},\ and\ \citenamefont {Drescher}}]{hartmann2019emergence}%
  \BibitemOpen
  \bibfield  {author} {\bibinfo {author} {\bibfnamefont {R.}~\bibnamefont {Hartmann}}, \bibinfo {author} {\bibfnamefont {P.}~\bibnamefont {Singh}}, \bibinfo {author} {\bibfnamefont {P.}~\bibnamefont {Pearce}}, \bibinfo {author} {\bibfnamefont {R.}~\bibnamefont {Mok}}, \bibinfo {author} {\bibfnamefont {B.}~\bibnamefont {Song}}, \bibinfo {author} {\bibfnamefont {F.}~\bibnamefont {D{\'\i}az-Pascual}}, \bibinfo {author} {\bibfnamefont {J.}~\bibnamefont {Dunkel}}, \ and\ \bibinfo {author} {\bibfnamefont {K.}~\bibnamefont {Drescher}},\ }\href@noop {} {\bibfield  {journal} {\bibinfo  {journal} {Nat. Phys.}\ }\textbf {\bibinfo {volume} {15}},\ \bibinfo {pages} {251} (\bibinfo {year} {2019})}\BibitemShut {NoStop}%
\bibitem [{\citenamefont {Qin}\ \emph {et~al.}(2020)\citenamefont {Qin}, \citenamefont {Fei}, \citenamefont {Bridges}, \citenamefont {Mashruwala}, \citenamefont {Stone}, \citenamefont {Wingreen},\ and\ \citenamefont {Bassler}}]{Qin2020}%
  \BibitemOpen
  \bibfield  {author} {\bibinfo {author} {\bibfnamefont {B.}~\bibnamefont {Qin}}, \bibinfo {author} {\bibfnamefont {C.}~\bibnamefont {Fei}}, \bibinfo {author} {\bibfnamefont {A.~A.}\ \bibnamefont {Bridges}}, \bibinfo {author} {\bibfnamefont {A.~A.}\ \bibnamefont {Mashruwala}}, \bibinfo {author} {\bibfnamefont {H.~A.}\ \bibnamefont {Stone}}, \bibinfo {author} {\bibfnamefont {N.~S.}\ \bibnamefont {Wingreen}}, \ and\ \bibinfo {author} {\bibfnamefont {B.~L.}\ \bibnamefont {Bassler}},\ }\href {\doibase 10.1126/science.abb8501} {\bibfield  {journal} {\bibinfo  {journal} {Science}\ } (\bibinfo {year} {2020}),\ 10.1126/science.abb8501}\BibitemShut {NoStop}%
\bibitem [{\citenamefont {Nijjer}\ \emph {et~al.}(2021)\citenamefont {Nijjer}, \citenamefont {Li}, \citenamefont {Zhang}, \citenamefont {Lu}, \citenamefont {Zhang},\ and\ \citenamefont {Yan}}]{nijjer2021mechanical}%
  \BibitemOpen
  \bibfield  {author} {\bibinfo {author} {\bibfnamefont {J.}~\bibnamefont {Nijjer}}, \bibinfo {author} {\bibfnamefont {C.}~\bibnamefont {Li}}, \bibinfo {author} {\bibfnamefont {Q.}~\bibnamefont {Zhang}}, \bibinfo {author} {\bibfnamefont {H.}~\bibnamefont {Lu}}, \bibinfo {author} {\bibfnamefont {S.}~\bibnamefont {Zhang}}, \ and\ \bibinfo {author} {\bibfnamefont {J.}~\bibnamefont {Yan}},\ }\href@noop {} {\bibfield  {journal} {\bibinfo  {journal} {Nat. Commun.}\ }\textbf {\bibinfo {volume} {12}},\ \bibinfo {pages} {6632} (\bibinfo {year} {2021})}\BibitemShut {NoStop}%
\bibitem [{\citenamefont {Nijjer}\ \emph {et~al.}(2023)\citenamefont {Nijjer}, \citenamefont {Li}, \citenamefont {Kothari}, \citenamefont {Henzel}, \citenamefont {Zhang}, \citenamefont {Tai}, \citenamefont {Zhou}, \citenamefont {Cohen}, \citenamefont {Zhang},\ and\ \citenamefont {Yan}}]{nijjer2023biofilms}%
  \BibitemOpen
  \bibfield  {author} {\bibinfo {author} {\bibfnamefont {J.}~\bibnamefont {Nijjer}}, \bibinfo {author} {\bibfnamefont {C.}~\bibnamefont {Li}}, \bibinfo {author} {\bibfnamefont {M.}~\bibnamefont {Kothari}}, \bibinfo {author} {\bibfnamefont {T.}~\bibnamefont {Henzel}}, \bibinfo {author} {\bibfnamefont {Q.}~\bibnamefont {Zhang}}, \bibinfo {author} {\bibfnamefont {J.-S.}\ \bibnamefont {Tai}}, \bibinfo {author} {\bibfnamefont {S.}~\bibnamefont {Zhou}}, \bibinfo {author} {\bibfnamefont {T.}~\bibnamefont {Cohen}}, \bibinfo {author} {\bibfnamefont {S.}~\bibnamefont {Zhang}}, \ and\ \bibinfo {author} {\bibfnamefont {J.}~\bibnamefont {Yan}},\ }\href@noop {} {\bibfield  {journal} {\bibinfo  {journal} {Nat. Phys.}\ ,\ \bibinfo {pages} {1}} (\bibinfo {year} {2023})}\BibitemShut {NoStop}%
\bibitem [{\citenamefont {Ford}\ \emph {et~al.}(2024)\citenamefont {Ford}, \citenamefont {Celora}, \citenamefont {Westbrook}, \citenamefont {Dalwadi}, \citenamefont {Walker}, \citenamefont {Baumann}, \citenamefont {Weijer}, \citenamefont {Pearce},\ and\ \citenamefont {Chubb}}]{ford2024pattern}%
  \BibitemOpen
  \bibfield  {author} {\bibinfo {author} {\bibfnamefont {H.}~\bibnamefont {Ford}}, \bibinfo {author} {\bibfnamefont {G.}~\bibnamefont {Celora}}, \bibinfo {author} {\bibfnamefont {E.}~\bibnamefont {Westbrook}}, \bibinfo {author} {\bibfnamefont {M.}~\bibnamefont {Dalwadi}}, \bibinfo {author} {\bibfnamefont {B.}~\bibnamefont {Walker}}, \bibinfo {author} {\bibfnamefont {H.}~\bibnamefont {Baumann}}, \bibinfo {author} {\bibfnamefont {C.}~\bibnamefont {Weijer}}, \bibinfo {author} {\bibfnamefont {P.}~\bibnamefont {Pearce}}, \ and\ \bibinfo {author} {\bibfnamefont {J.}~\bibnamefont {Chubb}},\ }\href@noop {} {\bibfield  {journal} {\bibinfo  {journal} {bioRxiv}\ ,\ \bibinfo {pages} {2024}} (\bibinfo {year} {2024})}\BibitemShut {NoStop}%
\bibitem [{\citenamefont {Brugu{\'e}s}\ and\ \citenamefont {Needleman}(2014)}]{brugues2014physical}%
  \BibitemOpen
  \bibfield  {author} {\bibinfo {author} {\bibfnamefont {J.}~\bibnamefont {Brugu{\'e}s}}\ and\ \bibinfo {author} {\bibfnamefont {D.}~\bibnamefont {Needleman}},\ }\href@noop {} {\bibfield  {journal} {\bibinfo  {journal} {Proc. Nat. Acad. Sci. U.S.A.}\ }\textbf {\bibinfo {volume} {111}},\ \bibinfo {pages} {18496} (\bibinfo {year} {2014})}\BibitemShut {NoStop}%
\bibitem [{\citenamefont {Oriola}\ \emph {et~al.}(2018)\citenamefont {Oriola}, \citenamefont {Needleman},\ and\ \citenamefont {Brugu{\'e}s}}]{oriola2018physics}%
  \BibitemOpen
  \bibfield  {author} {\bibinfo {author} {\bibfnamefont {D.}~\bibnamefont {Oriola}}, \bibinfo {author} {\bibfnamefont {D.}~\bibnamefont {Needleman}}, \ and\ \bibinfo {author} {\bibfnamefont {J.}~\bibnamefont {Brugu{\'e}s}},\ }\href@noop {} {\bibfield  {journal} {\bibinfo  {journal} {Annu. Rev. Biophys.}\ }\textbf {\bibinfo {volume} {47}},\ \bibinfo {pages} {655} (\bibinfo {year} {2018})}\BibitemShut {NoStop}%
\bibitem [{\citenamefont {Oriola}\ \emph {et~al.}(2020)\citenamefont {Oriola}, \citenamefont {J{\"u}licher},\ and\ \citenamefont {Brugu{\'e}s}}]{oriola2020active}%
  \BibitemOpen
  \bibfield  {author} {\bibinfo {author} {\bibfnamefont {D.}~\bibnamefont {Oriola}}, \bibinfo {author} {\bibfnamefont {F.}~\bibnamefont {J{\"u}licher}}, \ and\ \bibinfo {author} {\bibfnamefont {J.}~\bibnamefont {Brugu{\'e}s}},\ }\href@noop {} {\bibfield  {journal} {\bibinfo  {journal} {Proc. Nat. Acad. Sci. U.S.A.}\ }\textbf {\bibinfo {volume} {117}},\ \bibinfo {pages} {16154} (\bibinfo {year} {2020})}\BibitemShut {NoStop}%
\bibitem [{\citenamefont {Keber}\ \emph {et~al.}(2014)\citenamefont {Keber}, \citenamefont {Loiseau}, \citenamefont {Sanchez}, \citenamefont {DeCamp}, \citenamefont {Giomi}, \citenamefont {Bowick}, \citenamefont {Marchetti}, \citenamefont {Dogic},\ and\ \citenamefont {Bausch}}]{keber2014topology}%
  \BibitemOpen
  \bibfield  {author} {\bibinfo {author} {\bibfnamefont {F.}~\bibnamefont {Keber}}, \bibinfo {author} {\bibfnamefont {E.}~\bibnamefont {Loiseau}}, \bibinfo {author} {\bibfnamefont {T.}~\bibnamefont {Sanchez}}, \bibinfo {author} {\bibfnamefont {S.}~\bibnamefont {DeCamp}}, \bibinfo {author} {\bibfnamefont {L.}~\bibnamefont {Giomi}}, \bibinfo {author} {\bibfnamefont {M.}~\bibnamefont {Bowick}}, \bibinfo {author} {\bibfnamefont {M.}~\bibnamefont {Marchetti}}, \bibinfo {author} {\bibfnamefont {Z.}~\bibnamefont {Dogic}}, \ and\ \bibinfo {author} {\bibfnamefont {A.}~\bibnamefont {Bausch}},\ }\href@noop {} {\bibfield  {journal} {\bibinfo  {journal} {Science}\ }\textbf {\bibinfo {volume} {345}},\ \bibinfo {pages} {1135} (\bibinfo {year} {2014})}\BibitemShut {NoStop}%
\bibitem [{\citenamefont {Adkins}\ \emph {et~al.}(2022)\citenamefont {Adkins}, \citenamefont {Kolvin}, \citenamefont {You}, \citenamefont {Witthaus}, \citenamefont {Marchetti},\ and\ \citenamefont {Dogic}}]{adkins2022dynamics}%
  \BibitemOpen
  \bibfield  {author} {\bibinfo {author} {\bibfnamefont {R.}~\bibnamefont {Adkins}}, \bibinfo {author} {\bibfnamefont {I.}~\bibnamefont {Kolvin}}, \bibinfo {author} {\bibfnamefont {Z.}~\bibnamefont {You}}, \bibinfo {author} {\bibfnamefont {S.}~\bibnamefont {Witthaus}}, \bibinfo {author} {\bibfnamefont {M.}~\bibnamefont {Marchetti}}, \ and\ \bibinfo {author} {\bibfnamefont {Z.}~\bibnamefont {Dogic}},\ }\href@noop {} {\bibfield  {journal} {\bibinfo  {journal} {Science}\ }\textbf {\bibinfo {volume} {377}},\ \bibinfo {pages} {768} (\bibinfo {year} {2022})}\BibitemShut {NoStop}%
\bibitem [{\citenamefont {Tayar}\ \emph {et~al.}(2023)\citenamefont {Tayar}, \citenamefont {Caballero}, \citenamefont {Anderberg}, \citenamefont {Saleh}, \citenamefont {Marchetti},\ and\ \citenamefont {Dogic}}]{tayar2023controlling}%
  \BibitemOpen
  \bibfield  {author} {\bibinfo {author} {\bibfnamefont {A.}~\bibnamefont {Tayar}}, \bibinfo {author} {\bibfnamefont {F.}~\bibnamefont {Caballero}}, \bibinfo {author} {\bibfnamefont {T.}~\bibnamefont {Anderberg}}, \bibinfo {author} {\bibfnamefont {O.}~\bibnamefont {Saleh}}, \bibinfo {author} {\bibfnamefont {M.}~\bibnamefont {Marchetti}}, \ and\ \bibinfo {author} {\bibfnamefont {Z.}~\bibnamefont {Dogic}},\ }\href@noop {} {\bibfield  {journal} {\bibinfo  {journal} {Nat. Mater.}\ }\textbf {\bibinfo {volume} {22}},\ \bibinfo {pages} {1401} (\bibinfo {year} {2023})}\BibitemShut {NoStop}%
\bibitem [{\citenamefont {Fu}\ \emph {et~al.}(2024)\citenamefont {Fu}, \citenamefont {Huang}, \citenamefont {van~der Tol}, \citenamefont {Su}, \citenamefont {Wang}, \citenamefont {Dey}, \citenamefont {Zijlstra}, \citenamefont {Fytas}, \citenamefont {Vantomme}, \citenamefont {Dankers} \emph {et~al.}}]{fu2024supramolecular}%
  \BibitemOpen
  \bibfield  {author} {\bibinfo {author} {\bibfnamefont {H.}~\bibnamefont {Fu}}, \bibinfo {author} {\bibfnamefont {J.}~\bibnamefont {Huang}}, \bibinfo {author} {\bibfnamefont {J.}~\bibnamefont {van~der Tol}}, \bibinfo {author} {\bibfnamefont {L.}~\bibnamefont {Su}}, \bibinfo {author} {\bibfnamefont {Y.}~\bibnamefont {Wang}}, \bibinfo {author} {\bibfnamefont {S.}~\bibnamefont {Dey}}, \bibinfo {author} {\bibfnamefont {P.}~\bibnamefont {Zijlstra}}, \bibinfo {author} {\bibfnamefont {G.}~\bibnamefont {Fytas}}, \bibinfo {author} {\bibfnamefont {G.}~\bibnamefont {Vantomme}}, \bibinfo {author} {\bibfnamefont {P.}~\bibnamefont {Dankers}},  \emph {et~al.},\ }\href@noop {} {\bibfield  {journal} {\bibinfo  {journal} {Nature}\ }\textbf {\bibinfo {volume} {626}},\ \bibinfo {pages} {1011} (\bibinfo {year} {2024})}\BibitemShut {NoStop}%
\bibitem [{\citenamefont {Wensink}\ \emph {et~al.}(2012)\citenamefont {Wensink}, \citenamefont {Dunkel}, \citenamefont {Heidenreich}, \citenamefont {Drescher}, \citenamefont {Goldstein}, \citenamefont {L{\"o}wen},\ and\ \citenamefont {Yeomans}}]{wensink2012meso}%
  \BibitemOpen
  \bibfield  {author} {\bibinfo {author} {\bibfnamefont {H.}~\bibnamefont {Wensink}}, \bibinfo {author} {\bibfnamefont {J.}~\bibnamefont {Dunkel}}, \bibinfo {author} {\bibfnamefont {S.}~\bibnamefont {Heidenreich}}, \bibinfo {author} {\bibfnamefont {K.}~\bibnamefont {Drescher}}, \bibinfo {author} {\bibfnamefont {R.}~\bibnamefont {Goldstein}}, \bibinfo {author} {\bibfnamefont {H.}~\bibnamefont {L{\"o}wen}}, \ and\ \bibinfo {author} {\bibfnamefont {J.}~\bibnamefont {Yeomans}},\ }\href@noop {} {\bibfield  {journal} {\bibinfo  {journal} {Proc. Nat. Acad. Sci. U.S.A.}\ }\textbf {\bibinfo {volume} {109}},\ \bibinfo {pages} {14308} (\bibinfo {year} {2012})}\BibitemShut {NoStop}%
\bibitem [{\citenamefont {Khuc~Trong}\ \emph {et~al.}(2015)\citenamefont {Khuc~Trong}, \citenamefont {Doerflinger}, \citenamefont {Dunkel}, \citenamefont {St~Johnston},\ and\ \citenamefont {Goldstein}}]{khuc2015cortical}%
  \BibitemOpen
  \bibfield  {author} {\bibinfo {author} {\bibfnamefont {P.}~\bibnamefont {Khuc~Trong}}, \bibinfo {author} {\bibfnamefont {H.}~\bibnamefont {Doerflinger}}, \bibinfo {author} {\bibfnamefont {J.}~\bibnamefont {Dunkel}}, \bibinfo {author} {\bibfnamefont {D.}~\bibnamefont {St~Johnston}}, \ and\ \bibinfo {author} {\bibfnamefont {R.}~\bibnamefont {Goldstein}},\ }\href@noop {} {\bibfield  {journal} {\bibinfo  {journal} {eLife}\ }\textbf {\bibinfo {volume} {4}},\ \bibinfo {pages} {e06088} (\bibinfo {year} {2015})}\BibitemShut {NoStop}%
\bibitem [{\citenamefont {Saintillan}\ \emph {et~al.}(2018)\citenamefont {Saintillan}, \citenamefont {Shelley},\ and\ \citenamefont {Zidovska}}]{saintillan2018extensile}%
  \BibitemOpen
  \bibfield  {author} {\bibinfo {author} {\bibfnamefont {D.}~\bibnamefont {Saintillan}}, \bibinfo {author} {\bibfnamefont {M.}~\bibnamefont {Shelley}}, \ and\ \bibinfo {author} {\bibfnamefont {A.}~\bibnamefont {Zidovska}},\ }\href@noop {} {\bibfield  {journal} {\bibinfo  {journal} {Proc. Nat. Acad. Sci. U.S.A.}\ }\textbf {\bibinfo {volume} {115}},\ \bibinfo {pages} {11442} (\bibinfo {year} {2018})}\BibitemShut {NoStop}%
\bibitem [{\citenamefont {Needleman}\ and\ \citenamefont {Shelley}(2019)}]{Needleman2019stormy}%
  \BibitemOpen
  \bibfield  {author} {\bibinfo {author} {\bibfnamefont {D.}~\bibnamefont {Needleman}}\ and\ \bibinfo {author} {\bibfnamefont {M.}~\bibnamefont {Shelley}},\ }\href@noop {} {\bibfield  {journal} {\bibinfo  {journal} {Phys. Today}\ }\textbf {\bibinfo {volume} {72}},\ \bibinfo {pages} {6} (\bibinfo {year} {2019})}\BibitemShut {NoStop}%
\bibitem [{\citenamefont {Khalil}\ \emph {et~al.}(2014)\citenamefont {Khalil}, \citenamefont {Mahmoudi}, \citenamefont {Abu-Dheir},\ and\ \citenamefont {Varanasi}}]{khalil2014active}%
  \BibitemOpen
  \bibfield  {author} {\bibinfo {author} {\bibfnamefont {K.}~\bibnamefont {Khalil}}, \bibinfo {author} {\bibfnamefont {S.}~\bibnamefont {Mahmoudi}}, \bibinfo {author} {\bibfnamefont {N.}~\bibnamefont {Abu-Dheir}}, \ and\ \bibinfo {author} {\bibfnamefont {K.}~\bibnamefont {Varanasi}},\ }\href@noop {} {\bibfield  {journal} {\bibinfo  {journal} {App. Phys. Lett.}\ }\textbf {\bibinfo {volume} {105}} (\bibinfo {year} {2014})}\BibitemShut {NoStop}%
\bibitem [{\citenamefont {Aggarwal}\ \emph {et~al.}(2023{\natexlab{a}})\citenamefont {Aggarwal}, \citenamefont {Kirkinis},\ and\ \citenamefont {De~La~Cruz}}]{aggarwal2023activity}%
  \BibitemOpen
  \bibfield  {author} {\bibinfo {author} {\bibfnamefont {A.}~\bibnamefont {Aggarwal}}, \bibinfo {author} {\bibfnamefont {E.}~\bibnamefont {Kirkinis}}, \ and\ \bibinfo {author} {\bibfnamefont {M.}~\bibnamefont {De~La~Cruz}},\ }\href@noop {} {\bibfield  {journal} {\bibinfo  {journal} {J. Fluid Mech.}\ }\textbf {\bibinfo {volume} {955}},\ \bibinfo {pages} {A10} (\bibinfo {year} {2023}{\natexlab{a}})}\BibitemShut {NoStop}%
\bibitem [{\citenamefont {Aggarwal}\ \emph {et~al.}(2023{\natexlab{b}})\citenamefont {Aggarwal}, \citenamefont {Kirkinis},\ and\ \citenamefont {Olvera de~la Cruz}}]{aggarwal2023thermocapillary}%
  \BibitemOpen
  \bibfield  {author} {\bibinfo {author} {\bibfnamefont {A.}~\bibnamefont {Aggarwal}}, \bibinfo {author} {\bibfnamefont {E.}~\bibnamefont {Kirkinis}}, \ and\ \bibinfo {author} {\bibfnamefont {M.}~\bibnamefont {Olvera de~la Cruz}},\ }\href@noop {} {\bibfield  {journal} {\bibinfo  {journal} {Phys. Rev. Lett.}\ }\textbf {\bibinfo {volume} {131}},\ \bibinfo {pages} {198201} (\bibinfo {year} {2023}{\natexlab{b}})}\BibitemShut {NoStop}%
\bibitem [{\citenamefont {Tjhung}\ \emph {et~al.}(2012)\citenamefont {Tjhung}, \citenamefont {Marenduzzo},\ and\ \citenamefont {Cates}}]{Tjhung2012}%
  \BibitemOpen
  \bibfield  {author} {\bibinfo {author} {\bibfnamefont {E.}~\bibnamefont {Tjhung}}, \bibinfo {author} {\bibfnamefont {D.}~\bibnamefont {Marenduzzo}}, \ and\ \bibinfo {author} {\bibfnamefont {M.}~\bibnamefont {Cates}},\ }\href@noop {} {\bibfield  {journal} {\bibinfo  {journal} {Proc. Natl. Acad. Sci. U. S. A.}\ }\textbf {\bibinfo {volume} {109}},\ \bibinfo {pages} {12381} (\bibinfo {year} {2012})}\BibitemShut {NoStop}%
\bibitem [{\citenamefont {Tjhung}\ \emph {et~al.}(2015)\citenamefont {Tjhung}, \citenamefont {Tiribocchi}, \citenamefont {Marenduzzo},\ and\ \citenamefont {Cates}}]{Tjhung2015}%
  \BibitemOpen
  \bibfield  {author} {\bibinfo {author} {\bibfnamefont {E.}~\bibnamefont {Tjhung}}, \bibinfo {author} {\bibfnamefont {A.}~\bibnamefont {Tiribocchi}}, \bibinfo {author} {\bibfnamefont {D.}~\bibnamefont {Marenduzzo}}, \ and\ \bibinfo {author} {\bibfnamefont {M.~E.}\ \bibnamefont {Cates}},\ }\href@noop {} {\bibfield  {journal} {\bibinfo  {journal} {Nature communications}\ }\textbf {\bibinfo {volume} {6}},\ \bibinfo {pages} {1} (\bibinfo {year} {2015})}\BibitemShut {NoStop}%
\bibitem [{\citenamefont {P{\'e}rez-Gonz{\'a}lez}\ \emph {et~al.}(2019)\citenamefont {P{\'e}rez-Gonz{\'a}lez}, \citenamefont {Alert}, \citenamefont {Blanch-Mercader}, \citenamefont {G{\'o}mez-Gonz{\'a}lez}, \citenamefont {Kolodziej}, \citenamefont {Bazellieres}, \citenamefont {Casademunt},\ and\ \citenamefont {Trepat}}]{perez2019active}%
  \BibitemOpen
  \bibfield  {author} {\bibinfo {author} {\bibfnamefont {C.}~\bibnamefont {P{\'e}rez-Gonz{\'a}lez}}, \bibinfo {author} {\bibfnamefont {R.}~\bibnamefont {Alert}}, \bibinfo {author} {\bibfnamefont {C.}~\bibnamefont {Blanch-Mercader}}, \bibinfo {author} {\bibfnamefont {M.}~\bibnamefont {G{\'o}mez-Gonz{\'a}lez}}, \bibinfo {author} {\bibfnamefont {T.}~\bibnamefont {Kolodziej}}, \bibinfo {author} {\bibfnamefont {E.}~\bibnamefont {Bazellieres}}, \bibinfo {author} {\bibfnamefont {J.}~\bibnamefont {Casademunt}}, \ and\ \bibinfo {author} {\bibfnamefont {X.}~\bibnamefont {Trepat}},\ }\href@noop {} {\bibfield  {journal} {\bibinfo  {journal} {Nat. Phys.}\ }\textbf {\bibinfo {volume} {15}},\ \bibinfo {pages} {79} (\bibinfo {year} {2019})}\BibitemShut {NoStop}%
\bibitem [{\citenamefont {Alert}\ and\ \citenamefont {Casademunt}(2018)}]{alert2018role}%
  \BibitemOpen
  \bibfield  {author} {\bibinfo {author} {\bibfnamefont {.}~\bibnamefont {Alert}, \bibfnamefont {R}}\ and\ \bibinfo {author} {\bibfnamefont {J.}~\bibnamefont {Casademunt}},\ }\href@noop {} {\bibfield  {journal} {\bibinfo  {journal} {Langmuir}\ }\textbf {\bibinfo {volume} {35}},\ \bibinfo {pages} {7571} (\bibinfo {year} {2018})}\BibitemShut {NoStop}%
\bibitem [{\citenamefont {Liese}\ \emph {et~al.}(2023)\citenamefont {Liese}, \citenamefont {Zhao}, \citenamefont {Weber},\ and\ \citenamefont {J{\"u}licher}}]{liese2023chemically}%
  \BibitemOpen
  \bibfield  {author} {\bibinfo {author} {\bibfnamefont {S.}~\bibnamefont {Liese}}, \bibinfo {author} {\bibfnamefont {X.}~\bibnamefont {Zhao}}, \bibinfo {author} {\bibfnamefont {C.}~\bibnamefont {Weber}}, \ and\ \bibinfo {author} {\bibfnamefont {F.}~\bibnamefont {J{\"u}licher}},\ }\href@noop {} {\bibfield  {journal} {\bibinfo  {journal} {arXiv preprint arXiv:2312.07239}\ } (\bibinfo {year} {2023})}\BibitemShut {NoStop}%
\bibitem [{\citenamefont {Ben~Amar}\ and\ \citenamefont {Cummings}(2001)}]{ben2001fingering}%
  \BibitemOpen
  \bibfield  {author} {\bibinfo {author} {\bibfnamefont {M.}~\bibnamefont {Ben~Amar}}\ and\ \bibinfo {author} {\bibfnamefont {L.~J.}\ \bibnamefont {Cummings}},\ }\href@noop {} {\bibfield  {journal} {\bibinfo  {journal} {Phys. Fluids}\ }\textbf {\bibinfo {volume} {13}},\ \bibinfo {pages} {1160} (\bibinfo {year} {2001})}\BibitemShut {NoStop}%
\bibitem [{\citenamefont {Joanny}\ and\ \citenamefont {Ramaswamy}(2012)}]{joanny2012drop}%
  \BibitemOpen
  \bibfield  {author} {\bibinfo {author} {\bibfnamefont {J.~F.}\ \bibnamefont {Joanny}}\ and\ \bibinfo {author} {\bibfnamefont {S.}~\bibnamefont {Ramaswamy}},\ }\href@noop {} {\bibfield  {journal} {\bibinfo  {journal} {J. Fluid Mech.}\ }\textbf {\bibinfo {volume} {705}},\ \bibinfo {pages} {46} (\bibinfo {year} {2012})}\BibitemShut {NoStop}%
\bibitem [{\citenamefont {Loisy}\ \emph {et~al.}(2019)\citenamefont {Loisy}, \citenamefont {Eggers},\ and\ \citenamefont {Liverpool}}]{Loisy2019PRL}%
  \BibitemOpen
  \bibfield  {author} {\bibinfo {author} {\bibfnamefont {A.}~\bibnamefont {Loisy}}, \bibinfo {author} {\bibfnamefont {J.}~\bibnamefont {Eggers}}, \ and\ \bibinfo {author} {\bibfnamefont {T.~B.}\ \bibnamefont {Liverpool}},\ }\href@noop {} {\bibfield  {journal} {\bibinfo  {journal} {Phys. Rev. Lett.}\ }\textbf {\bibinfo {volume} {123}},\ \bibinfo {pages} {248006} (\bibinfo {year} {2019})}\BibitemShut {NoStop}%
\bibitem [{\citenamefont {Trinschek}\ \emph {et~al.}(2020)\citenamefont {Trinschek}, \citenamefont {Stegemerten}, \citenamefont {John},\ and\ \citenamefont {Thiele}}]{trinschek2020thin}%
  \BibitemOpen
  \bibfield  {author} {\bibinfo {author} {\bibfnamefont {S.}~\bibnamefont {Trinschek}}, \bibinfo {author} {\bibfnamefont {F.}~\bibnamefont {Stegemerten}}, \bibinfo {author} {\bibfnamefont {K.}~\bibnamefont {John}}, \ and\ \bibinfo {author} {\bibfnamefont {U.}~\bibnamefont {Thiele}},\ }\href@noop {} {\bibfield  {journal} {\bibinfo  {journal} {Phys. Rev. E}\ }\textbf {\bibinfo {volume} {101}},\ \bibinfo {pages} {062802} (\bibinfo {year} {2020})}\BibitemShut {NoStop}%
\bibitem [{\citenamefont {Loisy}\ \emph {et~al.}(2020)\citenamefont {Loisy}, \citenamefont {Eggers},\ and\ \citenamefont {Liverpool}}]{Loisy2020}%
  \BibitemOpen
  \bibfield  {author} {\bibinfo {author} {\bibfnamefont {A.}~\bibnamefont {Loisy}}, \bibinfo {author} {\bibfnamefont {J.}~\bibnamefont {Eggers}}, \ and\ \bibinfo {author} {\bibfnamefont {T.~B.}\ \bibnamefont {Liverpool}},\ }\href@noop {} {\bibfield  {journal} {\bibinfo  {journal} {Soft Matter}\ }\textbf {\bibinfo {volume} {16}},\ \bibinfo {pages} {3106} (\bibinfo {year} {2020})}\BibitemShut {NoStop}%
\bibitem [{\citenamefont {Shankar}\ \emph {et~al.}(2022)\citenamefont {Shankar}, \citenamefont {Raju},\ and\ \citenamefont {Mahadevan}}]{shankar2022optimal}%
  \BibitemOpen
  \bibfield  {author} {\bibinfo {author} {\bibfnamefont {S.}~\bibnamefont {Shankar}}, \bibinfo {author} {\bibfnamefont {V.}~\bibnamefont {Raju}}, \ and\ \bibinfo {author} {\bibfnamefont {L.}~\bibnamefont {Mahadevan}},\ }\href@noop {} {\bibfield  {journal} {\bibinfo  {journal} {Proceedings of the National Academy of Sciences}\ }\textbf {\bibinfo {volume} {119}},\ \bibinfo {pages} {e2121985119} (\bibinfo {year} {2022})}\BibitemShut {NoStop}%
\bibitem [{\citenamefont {Ioratim-Uba}\ \emph {et~al.}(2022)\citenamefont {Ioratim-Uba}, \citenamefont {Loisy}, \citenamefont {Henkes},\ and\ \citenamefont {Liverpool}}]{ioratim2022nonlinear}%
  \BibitemOpen
  \bibfield  {author} {\bibinfo {author} {\bibfnamefont {A.}~\bibnamefont {Ioratim-Uba}}, \bibinfo {author} {\bibfnamefont {A.}~\bibnamefont {Loisy}}, \bibinfo {author} {\bibfnamefont {S.}~\bibnamefont {Henkes}}, \ and\ \bibinfo {author} {\bibfnamefont {T.~B.}\ \bibnamefont {Liverpool}},\ }\href@noop {} {\bibfield  {journal} {\bibinfo  {journal} {Soft Matter}\ }\textbf {\bibinfo {volume} {18}},\ \bibinfo {pages} {9008} (\bibinfo {year} {2022})}\BibitemShut {NoStop}%
\bibitem [{\citenamefont {Oron}\ \emph {et~al.}(1997)\citenamefont {Oron}, \citenamefont {Davis},\ and\ \citenamefont {Bankoff}}]{oron1997long}%
  \BibitemOpen
  \bibfield  {author} {\bibinfo {author} {\bibfnamefont {A.}~\bibnamefont {Oron}}, \bibinfo {author} {\bibfnamefont {S.~H.}\ \bibnamefont {Davis}}, \ and\ \bibinfo {author} {\bibfnamefont {S.~G.}\ \bibnamefont {Bankoff}},\ }\href@noop {} {\bibfield  {journal} {\bibinfo  {journal} {Rev. Mod. Phys.}\ }\textbf {\bibinfo {volume} {69}},\ \bibinfo {pages} {931} (\bibinfo {year} {1997})}\BibitemShut {NoStop}%
\bibitem [{\citenamefont {Craster}\ and\ \citenamefont {Matar}(2009)}]{craster2009dynamics}%
  \BibitemOpen
  \bibfield  {author} {\bibinfo {author} {\bibfnamefont {R.~V.}\ \bibnamefont {Craster}}\ and\ \bibinfo {author} {\bibfnamefont {O.~K.}\ \bibnamefont {Matar}},\ }\href@noop {} {\bibfield  {journal} {\bibinfo  {journal} {Rev. Mod. Phys.}\ }\textbf {\bibinfo {volume} {81}},\ \bibinfo {pages} {1131} (\bibinfo {year} {2009})}\BibitemShut {NoStop}%
\bibitem [{\citenamefont {Needleman}\ and\ \citenamefont {Dogic}(2017)}]{needleman2017active}%
  \BibitemOpen
  \bibfield  {author} {\bibinfo {author} {\bibfnamefont {D.}~\bibnamefont {Needleman}}\ and\ \bibinfo {author} {\bibfnamefont {Z.}~\bibnamefont {Dogic}},\ }\href@noop {} {\bibfield  {journal} {\bibinfo  {journal} {Nat. Rev. Mater.}\ }\textbf {\bibinfo {volume} {2}},\ \bibinfo {pages} {1} (\bibinfo {year} {2017})}\BibitemShut {NoStop}%
\bibitem [{\citenamefont {Lee}\ \emph {et~al.}(2021)\citenamefont {Lee}, \citenamefont {Leech}, \citenamefont {Rust}, \citenamefont {Das}, \citenamefont {McGorty}, \citenamefont {Ross},\ and\ \citenamefont {Robertson-Anderson}}]{lee2021myosin}%
  \BibitemOpen
  \bibfield  {author} {\bibinfo {author} {\bibfnamefont {G.}~\bibnamefont {Lee}}, \bibinfo {author} {\bibfnamefont {G.}~\bibnamefont {Leech}}, \bibinfo {author} {\bibfnamefont {M.}~\bibnamefont {Rust}}, \bibinfo {author} {\bibfnamefont {M.}~\bibnamefont {Das}}, \bibinfo {author} {\bibfnamefont {R.}~\bibnamefont {McGorty}}, \bibinfo {author} {\bibfnamefont {J.}~\bibnamefont {Ross}}, \ and\ \bibinfo {author} {\bibfnamefont {R.}~\bibnamefont {Robertson-Anderson}},\ }\href@noop {} {\bibfield  {journal} {\bibinfo  {journal} {Sci. Adv.}\ }\textbf {\bibinfo {volume} {7}},\ \bibinfo {pages} {eabe4334} (\bibinfo {year} {2021})}\BibitemShut {NoStop}%
\bibitem [{\citenamefont {Berezney}\ \emph {et~al.}(2022)\citenamefont {Berezney}, \citenamefont {Goode}, \citenamefont {Fraden},\ and\ \citenamefont {Dogic}}]{berezney2022extensile}%
  \BibitemOpen
  \bibfield  {author} {\bibinfo {author} {\bibfnamefont {J.}~\bibnamefont {Berezney}}, \bibinfo {author} {\bibfnamefont {B.}~\bibnamefont {Goode}}, \bibinfo {author} {\bibfnamefont {S.}~\bibnamefont {Fraden}}, \ and\ \bibinfo {author} {\bibfnamefont {Z.}~\bibnamefont {Dogic}},\ }\href@noop {} {\bibfield  {journal} {\bibinfo  {journal} {Proc. Nat. Acad. Sci. U.S.A.}\ }\textbf {\bibinfo {volume} {119}},\ \bibinfo {pages} {e2115895119} (\bibinfo {year} {2022})}\BibitemShut {NoStop}%
\bibitem [{\citenamefont {Stephen}\ and\ \citenamefont {Straley}(1974)}]{stephen1974}%
  \BibitemOpen
  \bibfield  {author} {\bibinfo {author} {\bibfnamefont {M.~J.}\ \bibnamefont {Stephen}}\ and\ \bibinfo {author} {\bibfnamefont {J.~P.}\ \bibnamefont {Straley}},\ }\href@noop {} {\bibfield  {journal} {\bibinfo  {journal} {Rev. Mod. Phys.}\ }\textbf {\bibinfo {volume} {46}},\ \bibinfo {pages} {617} (\bibinfo {year} {1974})}\BibitemShut {NoStop}%
\bibitem [{\citenamefont {De~Gennes}\ and\ \citenamefont {Prost}(1993)}]{de1993}%
  \BibitemOpen
  \bibfield  {author} {\bibinfo {author} {\bibfnamefont {P.~G.}\ \bibnamefont {De~Gennes}}\ and\ \bibinfo {author} {\bibfnamefont {J.}~\bibnamefont {Prost}},\ }\href@noop {} {\emph {\bibinfo {title} {The physics of liquid crystals}}},\ Vol.~\bibinfo {volume} {83}\ (\bibinfo  {publisher} {Oxford university press},\ \bibinfo {year} {1993})\BibitemShut {NoStop}%
\bibitem [{\citenamefont {Prost}\ \emph {et~al.}(2015)\citenamefont {Prost}, \citenamefont {J{\"u}licher},\ and\ \citenamefont {Joanny}}]{Prost2015}%
  \BibitemOpen
  \bibfield  {author} {\bibinfo {author} {\bibfnamefont {J.}~\bibnamefont {Prost}}, \bibinfo {author} {\bibfnamefont {F.}~\bibnamefont {J{\"u}licher}}, \ and\ \bibinfo {author} {\bibfnamefont {J.-F.}\ \bibnamefont {Joanny}},\ }\href@noop {} {\bibfield  {journal} {\bibinfo  {journal} {Nat. Phys.}\ }\textbf {\bibinfo {volume} {11}},\ \bibinfo {pages} {111} (\bibinfo {year} {2015})}\BibitemShut {NoStop}%
\bibitem [{\citenamefont {Giomi}\ and\ \citenamefont {DeSimone}(2014)}]{Giomi2014}%
  \BibitemOpen
  \bibfield  {author} {\bibinfo {author} {\bibfnamefont {L.}~\bibnamefont {Giomi}}\ and\ \bibinfo {author} {\bibfnamefont {A.}~\bibnamefont {DeSimone}},\ }\href@noop {} {\bibfield  {journal} {\bibinfo  {journal} {Phys. Rev. Lett.}\ }\textbf {\bibinfo {volume} {112}},\ \bibinfo {pages} {147802} (\bibinfo {year} {2014})}\BibitemShut {NoStop}%
\bibitem [{\citenamefont {Alert}(2022)}]{alert2022fingering}%
  \BibitemOpen
  \bibfield  {author} {\bibinfo {author} {\bibfnamefont {R.}~\bibnamefont {Alert}},\ }\href@noop {} {\bibfield  {journal} {\bibinfo  {journal} {J. Phys. A: Mathematical and Theoretical}\ }\textbf {\bibinfo {volume} {55}},\ \bibinfo {pages} {234009} (\bibinfo {year} {2022})}\BibitemShut {NoStop}%
\end{thebibliography}%

\clearpage
\newpage
\setcounter{figure}{0}
\setcounter{table}{0}
\setcounter{equation}{0}
\renewcommand\theequation{S\arabic{equation}}
\renewcommand\thefigure{{S\arabic{figure}}}
\renewcommand\thetable{{S\arabic{table}}}
\renewcommand{\eqref}[1]{\textup{{\normalfont~(\ref{#1}}\normalfont)}}

\section{Supplementary Information}

\noindent \textbf{Non-dimensionalization}. To perform two-dimensional (2D) numerical simulations of the system of equations presented in the main text, we first nondimensionalize it using the characteristic scales also deduced in the main text. To this end, we introduce the following dimensionless variables for the position vector, time, velocity, and pressure fields, respectively,
\begin{equation}
\tilde{\boldm{x}} = \frac{\boldm{x}}{R}, \quad \tilde{t} = \frac{t}{\mu R/\gamma}, \quad \tilde{\boldm{u}} = \frac{\boldm{u}}{\gamma/\mu}, \quad \tilde{\Pi} = \frac{\Pi}{\gamma/R},    
\end{equation}
where tildes denote dimensionless variables here in the Supplementary Information. Using the above characteristic scales, the dimensionless volume and momentum conservation equations read, respectively:
\begin{subequations}\label{eq:dimless_continuity_momentum}
\begin{align}
\bnabla \bcdot \tilde{\boldm{u}} = 0, \, \, \, \boldm{0} = \bnabla \bcdot \tilde{\boldm{\sigma}},
\end{align}
\end{subequations}
where $\tilde{\boldm{\sigma}} = -\tilde{\Pi} \mathbfsf{I} + \bnabla \tilde{\boldm{u}} + (\bnabla \tilde{\boldm{u}})^{\textrm{T}} - \Ca_{\alpha} \boldm{p} \boldm{p}$ is the dimensionless stress tensor, and $\Ca_{\alpha} \equiv \alpha/(\gamma/R)$ is the active Capillary number. Equations~\eqref{eq:dimless_continuity_momentum} together with the equation for the instantaneous orientation field,
\begin{equation}
    \bnabla^2 \boldm{p} = \boldm{0},
\end{equation}
are a closed system of dimensionless equations that describe the morphodynamics of the surface-attached active drop. 

At the liquid-air interface of the drop we impose a kinematic condition specifying that the velocity of the interface is equal to the velocity of the fluid, precluding mass transfer across the interface, the stress balance, and the orientation of the units:
\begin{subequations}\label{eq:dimless_kinematic_surfacebalance}
\begin{align}
& \textrm{Fluid:} \nonumber & \\
& (\partial_{\tilde{t}} \tilde{\boldm{r}}_{\textrm{i}} - \tilde{\boldm{u}})\bcdot \hat{\boldm{n}} = 0, \, \, \tilde{\boldm{\sigma}} \bcdot \hat{\boldm{n}} = - \tilde{\mathcal{C}}\hat{\boldm{n}}
 \, \, \,  \text{at} \, \, \,  \tilde{\boldm{r}} = \tilde{\boldm{r}}_{\textrm{i}}, & \\
& \textrm{Active units:} \, \, 
\boldm{p} = \mathbfsf{R}_{\text{i}} \bcdot \hat{\boldm{t}} \, \,  \textrm{at} \, \, \tilde{\boldm{r}} = \tilde{\boldm{r}}_{\textrm{i}},
\end{align}
\end{subequations}
where $\tilde{\boldm{r}}_{\textrm{i}}$ is the dimensionless position of the interface, $\hat{\boldm{n}}$ and $\hat{\boldm{t}}$ are the unit normal and tangential vectors to the interface, respectively, and $\tilde{\mathcal{C}} \equiv \bnabla \bcdot \hat{\boldm{n}}$ is twice the dimensionless mean curvature of the liquid-air interface. Here, $\mathbfsf{R}_{\textrm{i}}$ is the rotation matrix prescribing the angle $\theta_{\textrm{i}}$ that the orientation of the units form with the tangent vector to the interface. For simplicity, we follow Refs.~\citep{Loisy2019PRL,Loisy2020} and consider only quarter turns in the orientation angle, i.e., $\theta_{\textrm{i}} = w_{\textrm{i}} \pi/2$, where $w \in \mathbb{Z}$ is the winding number (Fig.~\ref{fig:figure1}a). 

At the solid substrate, we impose no-permeation and no-slip conditions for the velocity field, i.e., $\tilde{\boldm{u}} = \boldm{0}$ at $\tilde{y} = 0$, and we prescribe the orientation of the active units with respect to the tangent vector to the substrate, equivalently to the condition at the interface, i.e., $\boldm{p} = \mathbfsf{R}_{\textrm{s}} \bcdot \hat{\boldm{e}}_x$, where $\mathbfsf{R}_{\textrm{s}}$ is the corresponding rotation matrix prescribing the angle $\theta_{\textrm{s}}$ at the substrate. For simplicity, we also assume that the orientation angle with the substrate only changes in quarter turns, and thus it is specified by an additional winding number $w_{\textrm{s}}$. 

As initial conditions, we consider the shape of the drop is semicircular and the fluid in the drop is at rest, i.e., $\tilde{\boldm{u}}(\tilde{\boldm{r}},\tilde{t}=0) = \boldm{0}$ and $\tilde{\Pi}(\tilde{\boldm{r}},\tilde{t} = 0) = 1$.\\

\noindent \textbf{Numerical simulations}. We carry out numerical simulations of Eqs.~\eqref{eq:dimless_continuity_momentum}-\eqref{eq:dimless_kinematic_surfacebalance} using the finite-element method. To this end, all the dimensionless equations are written in weak form by means of the corresponding integral scalar product, defined in terms of test functions for the pressure $\tilde{\Pi}(\tilde{\boldm{r}},\tilde{t})$, the velocity $\tilde{\boldm{u}}(\tilde{\boldm{r}},\tilde{t})$, and the orientation of active units $\tilde{\boldm{p}}(\tilde{\boldm{r}},\tilde{t})$. By using Green identities we obtain an integral bilinear system of equations for the set of variables and their corresponding test functions:
\begin{subequations}\label{eq:weak}
\begin{align}
& \int_{\tilde{A}} \text{d}\tilde{A} \, \bnabla \boldm{p}:\bnabla \tilde{\boldm{\phi}}_{p} = 0, &\\
& \int_{\tilde{A}} \text{d}\tilde{A}\, \tilde{\phi}_{\Pi} (\bnabla \cdot \tilde{\boldm{u}}) = 0, \label{eq:weak_orientation} &\\
& \int_{\tilde{A}} \text{d}\tilde{A} \, \tilde{\boldm{\sigma}}:\bnabla \tilde{\boldm{\phi}}_{u} - \int_{\tilde{\ell}} \textrm{d}\tilde{\ell} \, \tilde{\boldm{\sigma}} \bcdot  \hat{\boldm{n}} \bcdot \tilde{\boldm{\phi}}_{\boldm{u}}  = 0, \label{eq:weak_stress}
\end{align}
\end{subequations}
where $\tilde{\phi}_{\boldm{p}}$, $\tilde{\phi}_{\Pi}$, and $\tilde{\phi}_{\boldm{u}}$ are the test functions for the orientation, pressure, and velocity fields, respectively. Here, $\tilde{A}$ is the dimensionless area of the drop and $\tilde{\ell}$ the dimensionless length of the boundaries, where $\textrm{d}\tilde{A}$ and $\textrm{d}\tilde{\ell}$ are the corresponding surface and line elements, respectively. In Eq.~\eqref{eq:weak_orientation} we impose the Dirichlet boundary conditions as pointwise constraints on the boundaries for $\boldm{p}$, specifying the angle of the active units with the substrate and deformable interface. Similarly, we impose the no-flux and no-slip boundary conditions for $\tilde{\boldm{u}}$ on the substrate. To impose the stress balance at the deformable interface, the second term of Eq.~\eqref{eq:weak_stress} is expressed as: $\int_{\tilde{\ell}} \text{d} \tilde{\ell} \, \bnabla_{\text{s}} \bcdot \tilde{\boldm{\phi}}_{\boldm{u}}$, where $\bnabla_{\text{s}} = (\mathbfsf{I}-\boldm{n}\boldm{n})\bcdot \bnabla$ is the surface gradient operator.

Equations~\eqref{eq:weak} are discretized using Taylor-Hood triangular elements for pressure and velocity and their corresponding test functions, ensuring numerical stability, and second-order Lagrange polynomials for the orientation field and its test function. To account for the deformation of the interface, we use the arbitrary Lagrangian-Eulerian technique, which allows us to track the interface imposing the kinematic boundary condition by prescribing the normal velocity of the mesh elements along the interface. In particular, the displacement of the mesh elements is computed by solving the Laplace equation for the displacement field, i.e., $\bnabla^2 \tilde{\boldm{q}} = \boldm{0}$. Regarding the time-stepping, we employ a 4th-order variable-step BDF method. The tolerance of the nonlinear method is always set below $10^{-6}$. The time-dependent solver was complemented with an automatic remeshing algorithm which generates a new mesh when the triangular elements become significantly distorted due to large deformations of the drop.
\end{document}

\linenumbers

%\preprint{}


\title{Universal self-similar van der Waals breakup of ultrathin liquid films}



\author{A. Mart\'inez-Calvo}

\author{A. Sevilla}
\affiliation{\'Area de Mec\'anica de Fluidos, Departamento de Ingenier\'ia T\'ermica
y de Fluidos, Universidad Carlos III de Madrid. Avda. de la Universidad 30, 28911,
Legan\'es, Madrid, Spain.}
\email[]{alejandro.sevilla@uc3m.es}

\date{\today}

%\maketitle

\section{Supplemental Material}

%In this Supplemental Material we show the details of the local structure of the flow extracted from numerical simulations of cases I, II and III. 

In this Supplemental Material we show a direct comparison of the exponential thinning regime for two different $Bq$ and $\Theta = 1$, obtained with the complete Stokes and Boussinesq-Scriven equations, and the one-dimensional model derived in~\cite{MartinezSevilla2018}, and the numerical integration of the same model by~\cite{Wee2020}.

\begin{figure}
   \centering
    \includegraphics[width =\textwidth]{aux_1D_comp_WWKB2020}
    \caption{Radius of the filament as a function of time obtained from our numerical simulations of the 1D model and the one performed by~\cite{Wee2020}, with the same values of parameters reported therein: $k = 0.7$, $\epsilon = 0.4$, $\Bou = 10^{-3}$ and $\Theta = 1$. The inset shows a direct comparison between the 1D model and the complete Stokes and Boussinesq-Scriven equations but for $\Bou = 3.4$, fitted with the slope given by the function $F(\Theta)$.}
    \label{fig:fig1_sup}
\end{figure}

%%%% BIBLIO if any %%%%

\subsection{Linear stability and the limit of highly viscous interface}
\subsubsection{Passive ambient: modified Rayleigh-Chandraskhar dispersion relation}
In the Stokes limit reads
\begin{equation}
\omega = \frac{(1-k^2)[\Bou (\Theta + 1)(F(k)(F(k)-2)-k^2)-2]}{4[1-F(k)^2+k^2+\Bou^2 \Theta(F(k)(F(k)-2)-k^2)] -2 \Bou [1+4F(k)(F(k)-1) k^2 (\Theta-3) \Theta]}   
\end{equation}
When surface viscous stresses dominate, $\Bou \gg 1$, the dispersion relation becomes
\begin{equation}
\omega = \frac{(1-k^2)(1+\Theta)}{4 \Theta}
\end{equation}
\subsubsection{Liquid-liquid configuration: modified Tomotika's dispersion relation}

\subsection{Comparison with the leading-order one-dimensional model}

\bibliographystyle{apsrev}
\bibliography{biblio}